\documentclass[reprint, amsmath, amssymb, aps, prb, floatfix, superscriptaddress]{revtex4-2}

\usepackage[english]{babel}

\usepackage{placeins}
\usepackage{amsmath}

\usepackage{graphicx}
\usepackage{subfigure}

\usepackage{braket}
\usepackage{dcolumn}
\usepackage{amsfonts}
\usepackage{amssymb}
\usepackage{url}
\usepackage[colorlinks=true, allcolors=blue]{hyperref}
\usepackage[table]{xcolor}
\usepackage{lipsum}
\usepackage{float}

\graphicspath{{./}}
\begin{document}
\title{Josephson Diode Effect for a Kitaev Ladder System}
% APS accepted-manuscript deposit notice (author version, not Version of Record).
\thanks{\textcopyright\ 2026 American Physical Society. This is the accepted manuscript of the following article: Cheng-Rong Xie, Hiroki Tsuchiura, and Manfred Sigrist, ``Josephson diode effect for a Kitaev ladder system,'' Phys. Rev. B \textbf{114}, 334501 (2026). The final published version is available from \url{https://doi.org/10.1103/tmb4-f9zt}. This manuscript version is posted with permission for non-commercial scholarly use.}
%Manuscript Title:\\with Forced Linebreak}% Force line breaks with \\

\author{Cheng-Rong Xie}
\affiliation{Department of Applied Physics, Tohoku University, Sendai 980-8579, Japan}
\email{xie.chengrong.s4@dc.tohoku.ac.jp}
\affiliation{Institute for Advanced Study, Shenzhen University, Shenzhen 518060, China}
\altaffiliation[Also at ]{Institute for Theoretical Physics, ETH Zurich, 8093 Zurich, Switzerland.}

\author{Hiroki Tsuchiura}
\affiliation{Department of Applied Physics, Tohoku University, Sendai 980-8579, Japan}

\author{Manfred Sigrist}
\affiliation{Institute for Theoretical Physics, ETH Zurich, 8093 Zurich, Switzerland.}

\begin{abstract}
We study the Josephson diode effect arising from geometric and phase-structural properties in a Kitaev-ladder Josephson junction composed of two parallel spinless $p$-wave chains coupled by interleg hopping $t_\perp$.
The junction is governed by two phases: the superconducting phase difference across the junction, $\theta$, and the interleg phase difference, $\phi$.
For $\phi\notin \{0, \pi\}\; (\mathrm{mod}\;2\pi)$, the effective time-reversal-like constraint is broken. Together with the absence of leg-exchange symmetry, this yields the condition for a Josephson diode response.
%By resolving transport into the two normal-state bands obtained by diagonalizing the leg-space Hamiltonian, we show that the leg phase acts as an effective phase difference for the same-band ($p_\nu/p_{\nu}$) tunneling, whereas, within the normal-band channel decomposition, the opposite-band ($p_\nu/p_{-\nu}$) tunneling phase is gauge-equivalent to the bare junction phase. The phase-shifted, non-sinusoidal same-band contribution can interfere with approximately unshifted Andreev contributions of different CPR structure, providing a minimal mechanism for unequal positive and negative critical currents.
A detailed analysis identifies different tunneling contributions to the Josephson current-phase relation: a shifted, non-sinusoidal contribution and an approximately unshifted, sinusoidal contribution. Their interference provides a mechanism for unequal positive and negative critical currents.   
The class-D Pfaffian invariant identifies the parameter regimes in which protected Majorana zero modes occur at the ladder ends. Bogoliubov-de Gennes calculations reveal a pronounced dome-like dependence of the diode efficiency $\eta$ on $t_\perp$ with $\eta\to0$ for $t_\perp\to0$ and for large $t_\perp$, with a maximum at intermediate coupling, tunable by $\phi$.
Within this minimal topological ladder model, the results clarify how geometry, symmetry breaking, and topology combine to enable a superconducting diode effect in a one-dimensional system.

\end{abstract}

\maketitle

\section{Introduction}
Superconducting circuits that can ``rectify’’ supercurrents without an external bias have attracted considerable attention since the experimental discovery of the \emph{superconducting diode effect} (SDE) \cite{Ando2020, Pal2022, Jeon2022, Wu2022}, and the theoretical understanding of such effects has been reviewed recently \cite{Shaffer2025Theories}.
In a Josephson junction, the diode effect is defined by an asymmetry of the critical currents in opposite directions, $I_c^+ \neq I_c^-$.
In the dc equilibrium regime, such nonreciprocity is forbidden whenever a current-reversing symmetry constrains the current-phase relation (CPR) to satisfy an antisymmetry condition of the form $I(\theta) = -I(-\theta)$ (up to a constant phase shift).
Therefore, achieving a diode effect requires breaking all such current-reversing symmetries, which in practice typically involves the simultaneous breaking of time-reversal symmetry and inversion (or inversion-like) symmetry~\cite{Tokura2018, Yokoyama2014_PRB, Daido2022_PRL, Giazotto2011_PRB}.
%The emergence of even-harmonic or phase-shifted components in the CPR signals the relaxation of these symmetry constraints and can give rise to asymmetric critical currents under appropriate conditions.

The SDE has been experimentally observed in non-centrosymmetric bulk superconductors subject to antisymmetric spin-orbit coupling and external magnetic fields, where the combined breaking of time-reversal symmetry (TRS) and inversion symmetry (IS) induces magneto-chiral anisotropy near the superconducting transition temperature~\cite{Nonreciprocalchargetransportintwo-dimensionalnoncentrosymmetricsuperconductors,Supercurrentdiodeeffectandfinite-momentumsuperconductors,Aphenomenologicaltheoryofsuperconductordiodes,Daido2022_PRL,Legg2022, Ilic2022, Ma2025}.  
%In such systems, nonlinear paraconductivity leads to directional asymmetry in the critical current \cite{Daido2022_PRL,Legg2022, Ilic2022, Ma2025}.

Beyond bulk materials, Josephson junctions provide a versatile platform for realizing diode-like behavior, where various symmetry-breaking mechanisms have been implemented, including geometric asymmetry \cite{Zazunov2024,Moll2025,Coraiola2024_FluxTunable} and external fields \cite{Turini2022,EnhancingtheJosephsondiodeeffectwithMajoranaboundstates}. In particular, interferometer-based Josephson junctions can generate higher harmonic contributions in the CPR through the coexistence and interference of multiple bound-state channels, leading to nonreciprocal Josephson currents\cite{Fukaya2022,Theoryofgiantdiodeeffectind-wavesuperconductorjunctions,Superconductingweaklinks,TheoryofthenonreciprocalJosephsoneffect,Souto2022}.

The interference between Majorana-induced contributions and conventional Andreev bound states (ABSs) provides a natural mechanism for nonreciprocity in topological Josephson junctions\cite{EnhancingtheJosephsondiodeeffectwithMajoranaboundstates, Mondal2025, Rosen2024_TI4pi}.
A central question is how geometry, symmetry breaking, and topology determine the effective pairing phases responsible for a diode response in minimal one-dimensional junctions.

\begin{figure}[t]
\centering
\includegraphics[width=0.45\textwidth]
{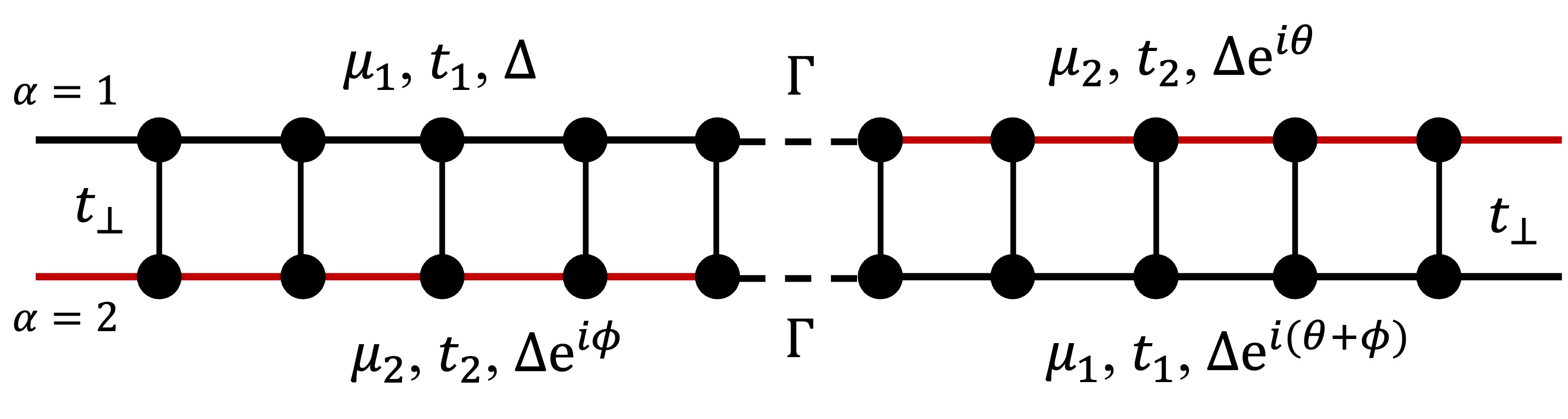}
\caption{The point-contact Josephson junction formed by two Kitaev ladders. $\phi$ is the phase difference between the legs, $\theta$ is the phase difference across the junction. $t_{\perp}$ is the interleg hopping amplitude. $\Gamma$ is the tunneling amplitude across the junction. $\mu_{1,2}$ and $t_{1,2}$ are the chemical potential and hopping amplitude of the two legs. Note that their assignments are interchanged across the junction.} 
\label{fig:ladder}
\end{figure}

In this work we analyze a \emph{Kitaev--ladder Josephson junction}
composed of two parallel wires with spinless $p$--wave Cooper pairing coupled by an
interchain hopping $t_{\perp}$ with a tunable phase
difference $\phi$ between the superconducting order parameters on the two legs (see Fig.~1).
This setup breaks the effective spinless time-reversal-like symmetry $\widetilde{\mathcal T}$ whenever $\phi \neq 0,\pi$ and
removes the leg-exchange symmetry that exchanges the two chains, while a weak
link imposes an additional tunable phase twist $\theta$ across the junction.
The combination of Pfaffian topological criteria and Bogoliubov--de~Gennes
numerics reveals that the ladder enters a topological phase with protected Majorana zero modes (MZMs) only within the class-D-nontrivial window of interleg coupling $t_\perp \neq 0 $.
Near the topological transition, low-energy phase-dispersive branches coexist with other Andreev contributions. A pronounced diode response can occur when the relevant contributions possess different gauge-invariant phases and nonidentical CPR structures or harmonic contents, their amplitudes are comparable, and no current-reversing symmetry applies to their sum.
The diode efficiency decreases as $t_{\perp} \to 0$, where the effective spinless time-reversal-like constraint is
effectively restored, and also vanishes for large $t_{\perp}$ due to the approximate recovery of current-reversal constraints\cite{Zinkl2022_PRR}.

The analysis identifies the symmetry and topological ingredients necessary for realizing superconducting rectification in ladder-type topological platforms.
The remainder of the paper is organized as follows:
Sec.~\ref{Topology of the Kitaev ladder} derives the Majorana number for the ladder,
Sec.~\ref{The quasi 1-d topological superconductor} presents its energy spectrum and phase diagram,
Sec.~\ref{Josephson Junction} analyzes the Josephson current analytically,
Sec.~\ref{Numerical Results} reports numerical simulations, and
Sec.~\ref{Conclusion} summarizes the findings.

%================

\section{Topology of the Kitaev ladder}\label{Topology of the Kitaev ladder}

In this section we introduce the Kitaev ladder model and briefly review
its topological properties~\cite{A.Yu.Kitaev2001,Maiellaro2018,Shen}, which provide the basis for the emergence of MZMs and the phase-dependent transport discussed in the following. Here we extend the standard Kitaev-ladder setting by allowing leg-dependent parameters (chemical potentials $\mu_1\neq\mu_2$ and hoppings $t_1\neq t_2$) together with a finite interleg superconducting phase difference $\phi$. These ingredients provide the symmetry structure required for the diode mechanism analyzed in Secs.~\ref{Josephson Junction} and \ref{Numerical Results}. Starting from the standard tight-binding model of a Kitaev chain~\cite{A.Yu.Kitaev2001}, 
\begin{align}
\begin{split}
H_{\text{chain}}=&-\mu \sum_{x=1}^{N}\left(c_x^{\dagger}c_{x}-\frac{1}{2}\right)-t\sum_{x=1}^{N-1}\left(c_x^{\dagger}c_{x+1}+c_{x+1}^{\dagger}c_{x}\right)\\
&+\sum_{x=1}^{N-1}\left(\Delta c_{x}c_{x+1}+\Delta^*c^{\dagger}_{x+1}c_x^{\dagger}\right),
\end{split}
\end{align}
with $t$ as the hopping amplitude, $\mu$ as the chemical potential, and $\Delta$ as the (real) pairing amplitude, we extend the Hamiltonian to a system with two coupled chains, the Kitaev ladder \cite{Maiellaro2018}
\begin{align}
\begin{split}
  H_{\text{lad}} =&\sum_{\alpha=1}^{2}\left[-\mu_{\alpha} \sum_{x=1}^{N}\left(c_{x,\alpha}^{\dagger}c_{x,\alpha}-\frac{1}{2}\right)\right.\\
  &\left.-\sum_{x=1}^{N-1}\left(t_{\alpha}c_{x,\alpha}^{\dagger}c_{x+1,\alpha}-\Delta_{\alpha}e^{i\phi_{\alpha}} c_{x,\alpha}c_{x+1,\alpha}+h.c.\right)\right]\\
  &-t_\perp\sum_{x=1}^{N}\left(c_{x,1}^{\dagger}c_{x,2}+h.c.\right).
\end{split}
\end{align}
Here $\alpha=1,2$ labels the two legs, with chemical potentials $\mu_\alpha$ and intraleg hopping amplitudes $t_\alpha$. The interleg hopping is $t_\perp$ (Fig.~\ref{fig:ladder}). We take equal pairing magnitudes $|\Delta_1|=|\Delta_2|=|\Delta|$, allowing different phases $\phi_\alpha$. With $\phi_1=0$ and $\phi_2=\phi$, the interleg phase difference is $\phi=\phi_2-\phi_1$. Assuming periodic boundary conditions, the ladder Hamiltonian can be written
in momentum space as
\begin{align}
H_{\text{lad}}
=\frac12 \sum_k
\Phi^\dagger(k)\,
\mathcal{H}_{\text{lad}}(k;\phi)\,
\Phi(k),
\end{align}
where
\begin{align}
\Phi(k)=
[c_1(k),c_2(k),c_1^\dagger(-k),c_2^\dagger(-k)]^{T}
\end{align}
is the Nambu spinor.
The corresponding Bogoliubov-de Gennes (BdG) Hamiltonian matrix reads
\begin{align}
\mathcal{H}_{\text{lad}}(k;\phi)=
\begin{bmatrix}
-\epsilon_{1,k} &-t_{\perp} & -\Delta_{1,k} & 0\\
-t_{\perp} &-\epsilon_{2,k} & 0 & -\Delta_{2,k}\\
-\Delta_{1,k}^{*} & 0 &\epsilon_{1,k} &t_{\perp}\\
 0 & -\Delta_{2,k}^{*} & t_{\perp} &\epsilon_{2,k}
\end{bmatrix}
\end{align}
where $\epsilon_{\alpha,k}=2t_{\alpha}\cos{k}+\mu_{\alpha}$ and $\Delta_{\alpha,k}=2i\Delta e^{-i\phi_{\alpha}}\sin{k}$. With the Fourier convention $c_{x,\alpha}=N^{-1/2}\sum_k e^{ikx}c_\alpha(k)$, the upper-right BdG block carries the complex-conjugate phase of the annihilation pairing. In the effective spinless model considered here, complex conjugation defines an antiunitary time-reversal-like operation
\begin{align}
\widetilde{\mathcal T}=\mathcal K,
\qquad
\widetilde{\mathcal T}^{\,2}=+1.
\end{align}
It transforms the BdG Hamiltonian according to
\begin{align}
\widetilde{\mathcal T}\,
\mathcal H_{\mathrm{lad}}(k;\phi)\,
\widetilde{\mathcal T}^{-1}
=
\mathcal H_{\mathrm{lad}}(-k;-\phi).
\end{align}
In this lattice gauge, $\widetilde{\mathcal T}=\mathcal K$ is a symmetry for $\phi=0$ or $\pi$ modulo $2\pi$, while generic $\phi\neq0,\pi$ breaks this effective time-reversal-like constraint. The particle-hole symmetry
\begin{align}
\Xi\,\mathcal H_{\mathrm{lad}}(k;\phi)\,\Xi^{-1}
=
-\mathcal H_{\mathrm{lad}}(-k;\phi),
\qquad
\Xi=\tau_x\otimes I_2\,\mathcal K,
\end{align}
remains present, where $\tau_x$ acts in the Nambu (particle-hole) space and $I_2$ is the identity in the two-dimensional leg space. Consequently, the generic finite-$\phi$ system belongs to symmetry class D. At $\phi=0$ or $\pi$, the additional $\widetilde{\mathcal T}$ symmetry enhances the symmetry class to BDI~\cite{Kitaev2009}.

For generic $\phi\neq0,\pi$, the class-D ladder is characterized by the $\mathbb Z_2$ Majorana number $\mathcal M$ (derived in Appendix~\ref{Derivation of the Majorana number}). At $\phi=0,\pi$, the system belongs to class BDI and admits an integer winding number. The invariant $\mathcal M$ records the parity of that winding number:
\begin{align}
\begin{split}
\mathcal{M}
=&\operatorname{sgn}[(2t_1+\mu_1)(2t_2+\mu_2)-t_{\perp}^2]\\
&\times\operatorname{sgn}[(2t_1-\mu_1)(2t_2-\mu_2)-t_{\perp}^2].    
\end{split}
\end{align}
The Majorana number $\mathcal M=\pm1$ is a class-D $\mathbb Z_2$ topological invariant. The value $\mathcal M=-1$ denotes the topologically nontrivial sector, whereas $\mathcal M=+1$ denotes the class-D-trivial sector. The sign of $\mathcal M$ specifies the parity of protected Majorana end modes. For the minimal ladder considered here, $\mathcal M=-1$ corresponds to one protected Majorana zero mode at each end. Assuming $\mu_\alpha, t_\alpha>0$, a topologically nontrivial state is found in the range,
\begin{align}\label{topological constraints0}
(2t_1-\mu_1)(2t_2-\mu_2)<t_{\perp}^2<(2t_1+\mu_1)(2t_2+\mu_2).
\end{align}

At $t_\perp=0$, the two legs are exactly decoupled and may be assigned individual class-D invariants $\mathcal M_1$ and $\mathcal M_2$. The invariant of the full ladder factorizes as
\begin{align}
\mathcal M=\mathcal M_1\mathcal M_2.
\end{align}
Consequently, the full class-D invariant is nontrivial only when exactly one leg is topological. If both legs are topological, $\mathcal M=+1$ because the full class-D invariant records only the parity of the number of Majorana channels. Nevertheless, the decoupled system retains one Majorana zero mode from each leg at a given end, protected by the exact block decoupling of the two topological legs. Once an interleg coupling is introduced, the fate of these two modes depends on the remaining symmetries. For generic $\phi\neq0,\pi$, where the system belongs to class D, the two modes are not protected by the total $\mathbb Z_2$ invariant and may hybridize and gap out. At $\phi=0,\pi$, the BDI winding number must instead be considered.

\begin{figure}[!htbp]
\centering
\includegraphics[width=40mm]
{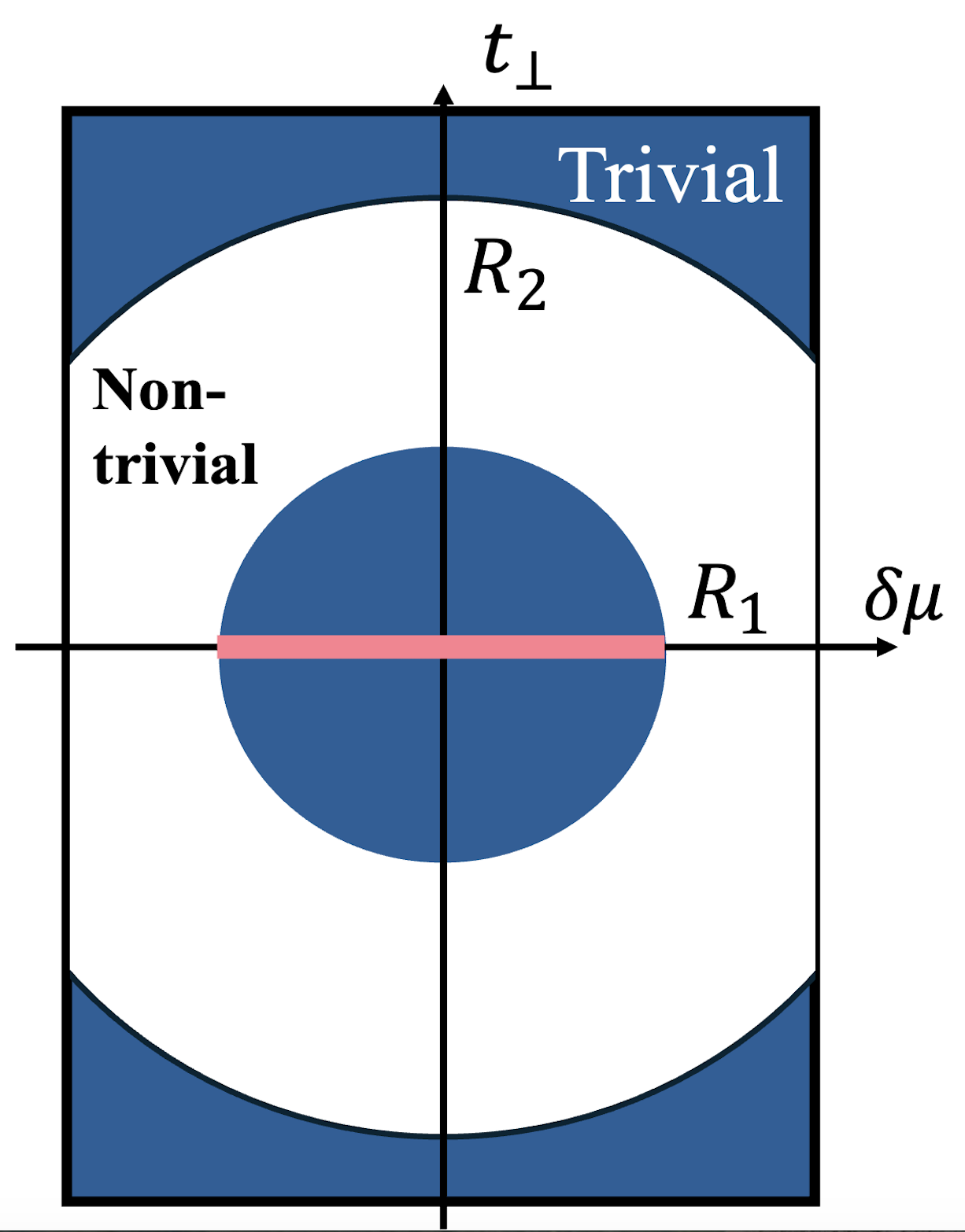}
\caption{The phase diagram of a Kitaev ladder as a function of the chemical potential difference $\delta \mu=\frac{\mu_2-\mu_1}{2}$ and perpendicular hopping term $t_{\perp}$. For $ t_1=t_2=t$ the radii of the two circles are given by $R_1=|2t-\frac{\mu_1+\mu_2}{2}|$ and $R_2=2t+\frac{\mu_1+\mu_2}{2}$. The white and blue regions refer to the total class-D invariant $\mathcal M$: white denotes $\mathcal M=-1$ (nontrivial) and blue denotes $\mathcal M=+1$ (trivial). Along the special line $t_\perp=0$, the two legs are decoupled and should also be characterized by their individual invariants $(\mathcal M_1,\mathcal M_2)$. The total class-D invariant satisfies $\mathcal M=\mathcal M_1\mathcal M_2$: it is nontrivial when exactly one leg is topological and trivial when the two legs have the same class-D topology. In particular, when both legs are topological, two decoupled Majorana channels are present although the total class-D invariant is $\mathcal M=+1$. The labels ``trivial'' and ``nontrivial'' refer to the class-D $\mathbb Z_2$ parity relevant for generic $\phi\neq0,\pi$. At $\phi=0,\pi$, the BDI winding number provides a finer classification.} \label{fig:topologyphasebreakinversion}
\end{figure}

We can simplify the phase diagram by setting $t_1=t_2=t$ such that the topologically nontrivial region lies between two concentric circles, corresponding to the white area displayed in Fig.\ref{fig:topologyphasebreakinversion} (see Appendix \ref{Topological criterion} for a more detailed discussion). 

In experimental implementations, the Kitaev chain can be emulated by proximitized semiconducting nanowires with strong spin–orbit coupling and Zeeman fields, which effectively realize one-dimensional topological superconductors hosting Majorana zero modes~\cite{Lutchyn2010, Oreg2010, Mourik2012, Das2012, Deng2016}. 
Coupling two such nanowires provides a potential route to engineer ladder-like structures, enabling the study of coupled topological superconducting channels~\cite{Klinovaja2014, Reeg2018}.

Although these systems are intrinsically spinful, their low-energy physics can often be mapped onto an effective spinless $p$-wave model~\cite{Alicea}. 
Therefore, the essential ingredients of our study—namely the interplay between topology (trivial versus nontrivial regimes) and effective phase structure—remain applicable beyond the minimal Kitaev-ladder model.

The Kitaev ladder provides a minimal low-energy model of a possible multi-terminal coupled-wire device. In such a device, an electrostatic barrier could tune the interleg hopping $t_\perp$, phase-biased superconducting contacts could control $\phi$, and local gates could set $\mu_1$ and $\mu_2$. Exchanging the gate settings across the junction would implement the leg-parameter arrangement shown in Fig.~\ref{fig:ladder}. Achieving these controls simultaneously remains experimentally challenging.

\section{Energy Spectrum and Topology}\label{The quasi 1-d topological superconductor}
We first study the energy spectrum of the Kitaev ladder with periodic boundary conditions. 
To elucidate the important topological features and, in particular, the emergence of a renormalized effective phase difference in the pairing potential, we additionally consider the continuum version of the Kitaev ladder with $t_{1}=t_{2}$ \cite{Kwon2004, Marra}. In this section, $H_{\mathrm{lad}}$ denotes the continuum counterpart of the lattice Hamiltonian introduced above. Since the absolute values of the pairing potentials do not affect the topological properties, we set them equal on both legs. The Hamiltonian is then given by
\begin{align}
\begin{split}
H_{\text{lad}}=&\sum_{\alpha=1}^2\int dx \left[\psi_{\alpha}^{\dagger}(x)(\frac{p_{x}^2}{2m}-\mu_{\alpha})\psi_{\alpha}(x)\right.\\
&+\left.\left(\frac{i\Delta e^{i\phi_{\alpha}}}{2\hbar}\psi_{\alpha}(x)p_{x}\psi_{\alpha}(x)+\mathrm{h.c.}\right) \right]\\
&-\int dx\left[\psi_{1}^{\dagger}(x)t_{\perp}\psi_{2}(x)+h.c.\right],
\end{split}
\label{Ham-lad-cont}
\end{align}
where $p_{x}=-i\hbar \partial_{x}$. With $t_{\perp}=0$, the two Kitaev chains are decoupled, and are individually topologically nontrivial for ${\mu_{\alpha}}>0$. To tackle the case of $t_{\perp}\neq 0$, we turn to momentum space. We choose different onsite potentials $\mu_1=\mu-\delta \mu$,  $\mu_2=\mu+\delta \mu$ for the two legs, with $\delta\mu\equiv(\mu_2-\mu_1)/2>0$ without loss of generality. With the gauge transformation $\psi_{\alpha}(x) \rightarrow \psi_{\alpha}(x)e^{-i\phi_{\alpha}/2}$, the Hamiltonian gets the form,
\begin{align}
\begin{split}
H_{\text{lad}}=&H_{0}+V\\
=&\int dx \Psi^{\dagger}(x) \left[\frac{p_{x}^2}{2m}-\mu+\delta \mu \, \sigma_{z}\right.\\
&\left.-t_{\perp}(\cos{\frac{\phi}{2}}\sigma_{x}+\sin{\frac{\phi}{2}}\sigma_{y})\right]\Psi(x)\\
&+\int dx\Bigl[\frac{i\Delta}{2\hbar}\Bigl(\Psi_1(x)p_{x}\Psi_1(x)\\
&\qquad\qquad\qquad+\Psi_2(x)p_{x}\Psi_2(x)\Bigr)+\mathrm{h.c.}\Bigr]
\end{split}
\end{align}
where we introduce the two-component spinor $\Psi(x)=\left[\psi_{1}(x), \psi_{2}(x)\right]^{T}$ in the leg space, with the Pauli matrices $ \sigma_{x,y,z}$ acting in the leg space. Note that $\Psi$ here does not include particle-hole (Nambu) degrees of freedom. With this choice of gauge, the phase difference between the legs is absorbed into the perpendicular hopping term $t_{\perp}$.  In momentum space, with $\Psi(k)=\left[\psi_{1}(k), \psi_{2}(k)\right]^{T}$, the Hamiltonian now reads
\begin{align}
\begin{split}
H_{\text{lad}}=&H_{0}+V\\
=&\int dk \Psi^{\dagger}(k) \left[\xi_{k}+\delta \mu \, \sigma_{z}\right.\\
&\left.\qquad\qquad\qquad -t_{\perp}(\cos{\frac{\phi}{2}}\sigma_{x}+\sin{\frac{\phi}{2}}\sigma_{y})\right]\Psi(k)\\
&-\frac12\int dk\Bigl[ik\Delta\Bigl(\Psi_1(k)\Psi_1(-k)\\
&\qquad\qquad\qquad+\Psi_2(k)\Psi_2(-k)\Bigr)+\mathrm{h.c.}\Bigr]
\end{split}
\end{align}
where $\xi_{k}=\frac{\hbar^2 k^2}{2m}-\mu$. We use $\psi_\alpha(x)=\int dk\,e^{ikx}\psi_\alpha(k)$ up to Fourier normalization, consistently with the lattice convention above. Here $H_0$ denotes the number-conserving part, including the kinetic energy, chemical potentials, and interleg hopping, whereas $V$ denotes the superconducting pairing. To illustrate the topological properties of the system, we diagonalize the normal-state leg-space Hamiltonian into two band sectors \cite{Shen, Geometry}. For this purpose, we first apply the unitary transformation to diagonalize the normal-state Hamiltonian ${H}_{0}$,
\begin{align}
\begin{pmatrix}
\psi_{1}(k)\\
\psi_{2}(k)
\end{pmatrix}
=
\begin{pmatrix}
\cos{\frac{\kappa}{2}} & e^{-\frac{i\phi}{2}}\sin{\frac{\kappa}{2}}\\
-e^{\frac{i\phi}{2}}\sin{\frac{\kappa}{2}} &\cos{\frac{\kappa}{2}}
\end{pmatrix}
\begin{pmatrix}
\psi_{+}(k)\\
\psi_{-}(k)
\end{pmatrix}\label{unitary transformation}
\end{align}
which yields, 
\begin{align}
H_{0}=\sum_{\nu=\pm }\int dk \psi_{\nu}^{\dagger}(k) \left[\xi_{k}+\nu \sqrt{\delta \mu^2+t_{\perp}^2}\right]\psi_{\nu}(k)
\end{align}
where 
\begin{align}
\cos \kappa = \frac{\delta \mu}{\sqrt{\delta \mu^2 + t_{\perp}^2}}, \quad
\sin \kappa  = \frac{t_{\perp}}{\sqrt{\delta \mu^2 + t_{\perp}^2}},
\end{align}
and the label $\nu=\pm$ denotes the two normal-state bands obtained after diagonalizing the leg-space Hamiltonian.
The pairing potential is divided into two parts, $V=V_{1}+V_{2}$,
where $V_{1}$ is the interband $p$-wave pairing potential, and $V_{2}$ is the intraband pairing potential:
\begin{align}
\begin{split}
V_{1}=&-\frac{1}{2}\int dk \Bigl[\Delta_{k,c}\Bigl(\psi_{+}(k)\psi_{-}(-k)\\
&\qquad\qquad\qquad+\psi_{-}(k)\psi_{+}(-k)\Bigr)+h.c.\Bigr],\\
V_{2}=&-\frac{1}{2}\sum_{\nu=\pm}\int dk\left[\Delta_{k,\nu}^{*}\,\psi_{\nu}(k)\psi_{\nu}(-k)+h.c.\right].
\end{split}
\end{align}
The factor $1/2$ avoids double counting in the integral over the full momentum range. Here $\Delta_{k,c}$ and $\Delta_{k,\nu}$ denote matrix elements of the upper-right BdG pairing block $\widehat{\Delta}_k$. With this Nambu convention, their annihilation-operator counterparts in $V$ have coefficients $-\Delta_{k,c}^{*}$ and $-\Delta_{k,\nu}^{*}$, respectively. The interband matrix element is real in the chosen band gauge.
The interband BdG pairing matrix element is
\begin{align}
\Delta_{k,c}
=
k\Delta\sin\frac{\phi}{2}
\frac{t_\perp}{\sqrt{\delta\mu^2+t_\perp^2}}
=
k\Delta\sin\frac{\phi}{2}\sin\kappa,\label{interband gap function}
\end{align}
and the intraband matrix element is
\begin{align}
\Delta_{k,\nu}= -i k \Delta e^{- i\nu \frac{\phi}{2}} (\cos \frac{\phi}{2} + i \nu \sin \frac{\phi}{2} \frac{\delta \mu}{\sqrt{\delta \mu^2+t_{\perp}^2}}).
\end{align}
In the Nambu representation, the pairing sector of the BdG Hamiltonian takes the form
\begin{align}
\widehat{\Delta}_k
=
\begin{pmatrix}
\Delta_{k,+} & \Delta_{k,c}\\[4pt]
\Delta_{k,c} & \Delta_{k,-}
\end{pmatrix},
\end{align}
where the diagonal entries originate from $V_2$ and the off-diagonal entries from $V_1$.
The Hamiltonian $H_c=H_0+V_1$ describes two normal-state bands coupled by an interband pairing potential.

After diagonalizing $H_0$ in the leg space, the pairing term is decomposed as $V=V_1+V_2$, where $V_1$ contains interband pairing between the two normal-state bands and $V_2$ contains intraband pairing within each normal-state band. The interband-paired Hamiltonian $ H_c$ is diagonalized by a Bogoliubov transformation. We denote the resulting transformed-branch index by $\lambda=\pm$ to distinguish it from the normal-band index $\nu=\pm$. Defining
\begin{align}
r\equiv\sqrt{\delta\mu^2+t_\perp^2},
\qquad
q_k\equiv\sqrt{\xi_k^2+\Delta_{k,c}^{\,2}},
\end{align}
the transformation reads
\begin{align}
\begin{pmatrix}
\psi_{+}(k)\\
\psi_{-}^{\dagger}(-k)
\end{pmatrix}
=
\begin{pmatrix}
\cos{\frac{\beta_k}{2}} & -\sin{\frac{\beta_k}{2}}\\[4pt]
\sin{\frac{\beta_k}{2}} &\cos{\frac{\beta_k}{2}}
\end{pmatrix}
\begin{pmatrix}
\chi_{+}(k)\\
\chi_{-}^{\dagger}(-k)
\end{pmatrix}\label{Bogoliubov transformation}
\end{align}
together with the particle-hole-conjugate transformation acting on $(\psi_-(k),\psi_+^\dagger(-k))^T$. The Bogoliubov angle is defined by
\begin{align}
\cos\beta_k=\frac{\xi_k}{q_k},
\qquad
\sin\beta_k=\frac{\Delta_{k,c}}{q_k}.
\end{align}
On the left-hand side, the signs attached to $\psi_\nu$ denote the original normal-state band index $\nu=\pm$. On the right-hand side, the signs attached to $\chi_\lambda$ denote the transformed-branch index $\lambda=\pm$. In the $\chi_\lambda$ basis,
\begin{align}
H_c
=
\sum_{\lambda=\pm}
\int dk\,
\chi_\lambda^\dagger(k)
\left(q_k+\lambda r\right)
\chi_\lambda(k).
\end{align}
Up to an additive constant, $H_c$ has the signed dispersions $\zeta_\lambda(k)=q_k+\lambda r$. Including $V_2$ yields the nonnegative quasiparticle energies $\epsilon_\lambda(k)\geq0$ of the complete BdG Hamiltonian.
Although the Bogoliubov transformation mixes particle and hole operators originating from opposite normal-state bands, the complete $4\times4$ BdG Hamiltonian can be diagonalized directly. Its spectrum can be organized into two positive-energy quasiparticle branches labeled by $\lambda=\pm$ and their particle-hole partners.

Introducing the common intraband-pairing magnitude
\begin{align}
D_k\equiv|\Delta_{k,+}|=|\Delta_{k,-}|,
\end{align}
the positive excitation energies are
\begin{align}
\epsilon_\lambda(k)
=
\sqrt{
D_k^2+
\left(r+\lambda q_k\right)^2
},
\qquad
\lambda=\pm.
\label{excitation spectrum}
\end{align}
The full particle-hole-symmetric BdG spectrum is then
\begin{align}
E_{\lambda,s}(k)
=
s\,\epsilon_\lambda(k),
\qquad
s=\pm,
\end{align}
where $s=+$ and $s=-$ give the positive and negative branches as discussed below.  For finite interleg coupling and generic $\phi\neq0,\pi$, the $\lambda=+$ branch cannot close its gap because $r+q_k>0$. The class-D bulk gap closing therefore occurs through the $\lambda=-$ branch. At the exactly decoupled point $t_\perp=0$, the leg-resolved topological invariants provide the more appropriate description, as discussed in Sec.~\ref{Topology of the Kitaev ladder} and Appendix~\ref{Topological criterion}. In the limit $t_\perp\gg|\delta\mu|$, the $\lambda=+$ branch is pushed to high energies and may be neglected in the low-energy description at zero temperature~\cite{Geometry,Alicea}.

\begin{figure}[!htbp]
\centering
\includegraphics[width=80mm]
{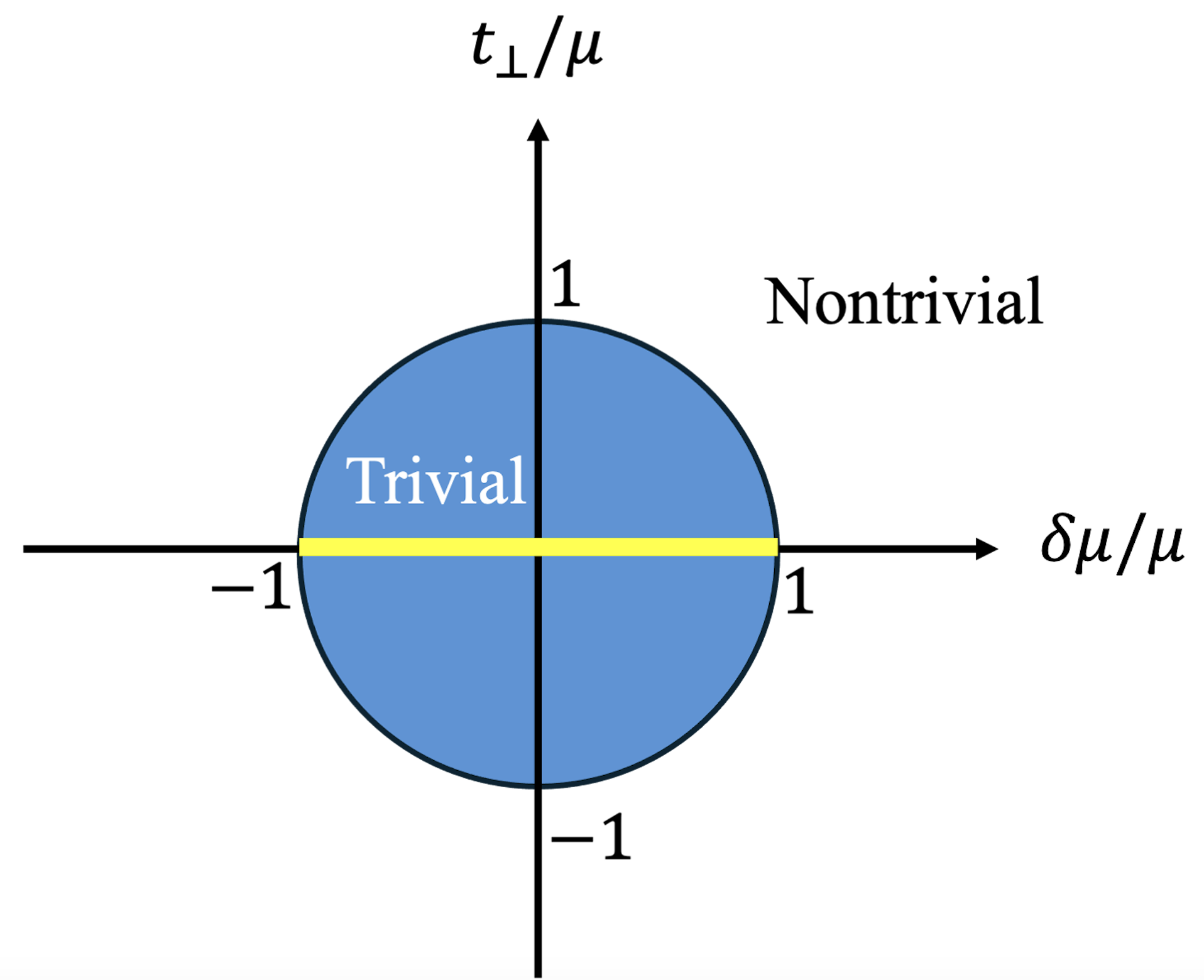}
\caption{The phase diagram of a Kitaev ladder as a function of the chemical potential difference $\delta \mu$ and perpendicular hopping term $t_{\perp}$. The topologically trivial and nontrivial phases are realized  at  $\sqrt{(\delta \mu/\mu)^2+(t_{\perp}/\mu)^2}<1$ and $\sqrt{(\delta \mu/\mu)^2+(t_{\perp}/\mu)^2}>1$, respectively, in the blue and white region. Along the $t_{\perp}=0$ axis, the two legs are decoupled and the total class-D invariant factorizes as $\mathcal M=\mathcal M_1\mathcal M_2$: it is nontrivial when exactly one leg is topological and trivial when both legs have the same topology. When both legs are individually topological, two decoupled Majorana channels are present although $\mathcal M=+1$. The labels ``trivial'' and ``nontrivial'' refer to the class-D $\mathbb Z_2$ parity relevant for generic $\phi\neq0,\pi$. At $\phi=0,\pi$, the BDI winding number provides a finer classification.} \label{The phase diagram of a Kitaev ladder}
\end{figure}

Because all $p$-wave pairing amplitudes vanish at $k=0$, the class-D bulk gap closes at $k=0$ when
\begin{align}
r=|\mu|.
\end{align}
It is therefore convenient to introduce the topological mass parameter
\begin{align}
m_{\mathrm{top}}
\equiv
r-|\mu|.
\end{align}
The generic finite-$\phi$ system is in the class-D nontrivial sector for $m_{\mathrm{top}}>0$ and in the class-D trivial sector for $m_{\mathrm{top}}<0$. In Fig. \ref{The phase diagram of a Kitaev ladder}, we plot the phase diagram of a Kitaev ladder. For $t_\perp=0$ and $\mu\pm\delta\mu>0$, the two decoupled continuum chains are individually in their topological regimes. The total class-D invariant of the full two-leg system is nevertheless the product of the two leg invariants and is therefore trivial under the class-D $\mathbb Z_2$ reduction when both legs are topological. The decoupled limit is more completely described by the pair of leg-resolved invariants.

\begin{figure}[!htbp]
\centering
\includegraphics[width=0.98\columnwidth]
{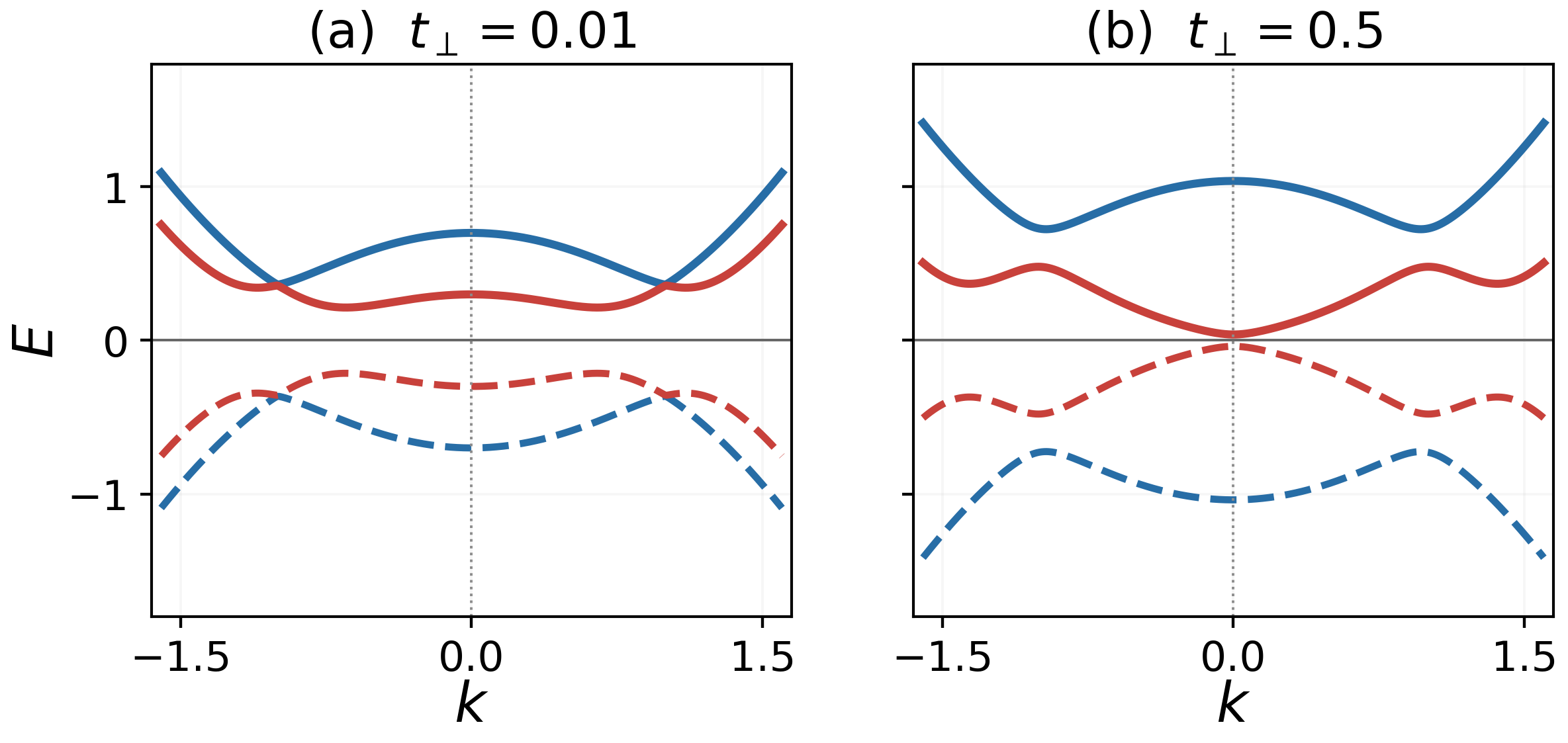}
\caption{The continuum BdG spectrum of the Kitaev ladder from Eq.~\eqref{excitation spectrum}. Blue and red indicate the $\lambda=+$ and $\lambda=-$ branches, while solid and dashed lines distinguish positive and negative energies. Parameters: $\mu=0.5$, $\delta\mu=0.2$, $\Delta=0.3$, $\phi=\pi/3$, $m=\hbar=1$. (a) $t_\perp=0.01$ (class-D trivial), (b) $t_\perp=0.5$ (class-D nontrivial). Both spectra have a finite gap at $k=0$.} \label{fig:2}
\end{figure}
In Fig.~\ref{fig:2}, we plot the quasiparticle spectrum obtained from
Eq.~\eqref{excitation spectrum} for two representative values of the
interleg hopping $t_{\perp}$. The blue and red curves correspond to
the two branches labeled by $\lambda=+$ and $\lambda=-$, respectively. For weak
interleg coupling [Fig.~\ref{fig:2}(a)], the two branches are only
weakly split, while for stronger interleg coupling
[Fig.~\ref{fig:2}(b)] the separation between them becomes more
pronounced. For the reduced interband-pairing Hamiltonian $H_c=H_0+V_1$, the avoided-crossing gap at $\xi_k=0$ is $2|\Delta_{k,c}|$. The full BdG excitation energies in Eq.~\eqref{excitation spectrum} also contain the intraband amplitude $D_k$, so the separation of their positive-energy branches is generally different.

We now analyze the magnitude and phase of the intraband pairing matrix elements $\Delta_{k,\nu}$ defined above. We factor out the common odd-parity structure $-ik\Delta$ and write
\begin{align}
\begin{split}
\Delta_{k,\nu}
=&
-ik\Delta\,A_\nu(\phi,\kappa),\\
A_\nu(\phi,\kappa)
=&
e^{-i\nu\phi/2}
\left(
\cos\frac{\phi}{2}
+
i\nu\cos\kappa\sin\frac{\phi}{2}
\right),
\end{split}
\end{align}
Writing $\cos(\phi/2)=(e^{i\phi/2}+e^{-i\phi/2})/2$ and $\sin(\phi/2)=(e^{i\phi/2}-e^{-i\phi/2})/(2i)$, together with $\cos\kappa=2\cos^2(\kappa/2)-1=1-2\sin^2(\kappa/2)$, one obtains the compact form
\begin{align}
A_\nu(\phi,\kappa)
=
\cos^2\frac{\kappa}{2}
+
\sin^2\frac{\kappa}{2}e^{-i\nu\phi}.
\end{align}
The effective phase is defined relative to the common factor $-ik\Delta$.

The magnitude is independent of $\nu$ and given by
\begin{align}
\mathcal A(\phi,\kappa)
\equiv
|A_\nu(\phi,\kappa)|
=
\sqrt{
1-\sin^2\kappa\sin^2\frac{\phi}{2}
}.
\end{align}
Consequently, the intraband pairing amplitude entering the BdG spectrum is
\begin{align}
D_k
\equiv
|\Delta_{k,\nu}|
=
|k\Delta|\,
\mathcal A(\phi,\kappa)
=
|k\Delta|
\sqrt{
1-\sin^2\kappa\sin^2\frac{\phi}{2}
}.
\label{intraband pairing potential}
\end{align}
This definition is consistent with the $D_k$ already used in the excitation spectrum, Eq.~\eqref{excitation spectrum}.

The effective intraband pairing phase is defined by
\begin{align}
\Delta_{k,\nu}
=
-ik\Delta\,
\mathcal A(\phi,\kappa)
e^{i\phi_{\nu,\mathrm{eff}}(\phi,\kappa)},
\end{align}
with
\begin{align}
\phi_{\nu,\mathrm{eff}}(\phi,\kappa)
=
\operatorname{Arg}
\left[
\cos^2\frac{\kappa}{2}
+
\sin^2\frac{\kappa}{2}e^{-i\nu\phi}
\right],
\label{effective phase of the intraband pairing gap functio}
\end{align}
where $\operatorname{Arg}$ takes its principal value in $(-\pi,\pi]$. Continuous phase tracking may differ by integer multiples of $2\pi$.
The effective phase is defined only when $\mathcal A(\phi,\kappa)\neq0$. At $\kappa=\pi/2$ and $\phi=\pi$ modulo $2\pi$, the intraband pairing amplitude vanishes and its phase is therefore undefined.

The parameter $\kappa$ satisfies
\begin{align}
\begin{split}
\cos\kappa=\frac{\delta\mu}{r},
\qquad
\sin\kappa=\frac{t_\perp}{r},\\
r=\sqrt{\delta\mu^2+t_\perp^2}.
\end{split}
\end{align}
For $t_\perp\ge0$ we choose $0\le\kappa\le\pi$. Changing $\delta\mu\to-\delta\mu$ maps $\kappa\to\pi-\kappa$. We demonstrate in Sec.~\ref{Josephson Junction} that the renormalization of intraband phases $\phi_{\nu,\mathrm{eff}}$ plays a crucial role for the Josephson diode effect.

We now examine two limiting cases.

\paragraph{Weak interleg coupling.}
For $t_\perp\ll|\delta\mu|$, we have $\kappa\to0$ when $\delta\mu>0$ and $\kappa\to\pi$ when $\delta\mu<0$. In both limits $\mathcal A\to1$, and
\begin{align}
\phi_{\nu,\mathrm{eff}}
\to
\begin{cases}
0, & \delta\mu>0,\\[4pt]
-\nu\phi, & \delta\mu<0,
\end{cases}
\qquad (\mathrm{mod}\ 2\pi).
\end{align}

\paragraph{Strong interleg coupling.}
For $t_\perp\gg|\delta\mu|$, we have $\kappa\to\pi/2$, so
\begin{align}
A_\nu
=
\frac{1+e^{-i\nu\phi}}{2}
=
\cos\frac{\phi}{2}\,e^{-i\nu\phi/2}.
\end{align}
For the principal interval $-\pi<\phi<\pi$,
\begin{align}
\mathcal A
=
\left|\cos\frac{\phi}{2}\right|,
\qquad
\phi_{\nu,\mathrm{eff}}
=
-\nu\frac{\phi}{2}.
\end{align}
At $\phi=\pi$, the intraband amplitude vanishes and the phase is undefined.

\section{Josephson Junction}\label{Josephson Junction}
% MODULE 3.2: Re-derive the gauge-invariant p_nu/p_nu and
% p_nu/p_-nu channel phases using the corrected phi_{nu,eff}.
Now we extend the setup to include a Josephson junction that separates two semi-infinite Kitaev ladders. The junction geometry is illustrated in Fig.~\ref{fig:ladder}. This allows us to explore the conditions for the appearance of a Josephson diode effect in such a system. We implement a weak-link model corresponding to a superconductor/insulator/superconductor (S/I/S) junction in the short-junction limit, described by the Hamiltonian \cite{Superconductor/normal-metal/superconductorjunctionoftopologicalsuperconductors,JosephsoneffectsbetweentheKitaevladdersuperconductors}
\begin{align}
H=H_{\text{SL}}+H_{\text{SR}}+H_{\Gamma},
\end{align}
with 
\begin{align}
\begin{split}
H_{\text{SL}} =&\sum_{\alpha=1}^{2}\left[-\mu_{\alpha} \sum_{x=-\infty}^{0}\left(c_{x,\alpha}^{\dagger}c_{x,\alpha}-\frac{1}{2}\right)\right.\\
  &\left.-\sum_{x=-\infty}^{-1}\left(t_{\alpha}c_{x,\alpha}^{\dagger}c_{x+1,\alpha}-\Delta c_{x,\alpha}c_{x+1,\alpha}+h.c.\right)\right]\\
  &-\sum_{x=-\infty}^{0}\left(t_\perp e^{-\frac{i\phi}{2}}c_{x,1}^{\dagger}c_{x,2}+h.c.\right),    
\end{split}
\end{align}

\begin{align}
\begin{split}
H_{\text{SR}} =&\sum_{\alpha=1}^{2}\left[-\mu_{3-\alpha} \sum_{x=1}^{\infty}\left(c_{x,\alpha}^{\dagger}c_{x,\alpha}-\frac{1}{2}\right)\right.\\
  &\left.-\sum_{x=1}^{\infty}\left(t_{3-\alpha}c_{x,\alpha}^{\dagger}c_{x+1,\alpha}-\Delta e^{i\theta} c_{x,\alpha}c_{x+1,\alpha}+h.c.\right)\right]\\
  &-\sum_{x=1}^{\infty}\left(t_\perp e^{-\frac{i\phi}{2}}c_{x,1}^{\dagger}c_{x,2}+h.c.\right),    
\end{split}
\end{align}
\begin{align}\label{tunneling}
H_{\Gamma}=\Gamma \sum_{\alpha=1}^{2} \left(c_{0,\alpha}^{\dagger}c_{1,\alpha}+h.c.\right).
\end{align}
The tunneling across the insulating weak link is included in $H_{\Gamma}$ with the tunneling amplitude $\Gamma$. %This corresponds to the short-junction limit where the insulating barrier is reduced to an effective point-contact coupling.

To better understand the properties of the diode effect, we consider again the continuum model. We introduce the four-component Nambu spinor $\Phi(x)=[\psi_1(x),\psi_2(x),\psi_1^\dagger(x),\psi_2^\dagger(x)]^T$, with the Pauli matrices $\tau_i$ acting in the particle-hole (Nambu) space and $\sigma_i$ in the two-dimensional leg space. In this basis the BdG Hamiltonian follows from Eq.~\eqref{Ham-lad-cont} by placing the normal-state single-particle matrix and its negative transpose in the electron and hole blocks, respectively, with pairing in the off-diagonal blocks. In the continuum formulation we realize the Josephson junction by a simple potential barrier at $x=0$ with $U(x)=U_0\delta(x)$, leading to the BdG Hamiltonian
\begin{align}
\begin{split}
\mathcal{H}(x)=&\left[\left(\frac{p_{x}^2}{2m}-\mu+U(x)\right)\sigma_{0}+\delta\mu(x)\sigma_{z}\right]\tau_{z}\\
&-t_{\perp}\cos\frac{\phi}{2}\,\sigma_{x}\,\tau_{z}
-t_{\perp}\sin\frac{\phi}{2}\,\sigma_{y}\,\tau_{0}\\
&-\frac{\Delta}{2\hbar}\Bigl\{p_{x},\,\sin\theta(x)\,\tau_{x}-\cos\theta(x)\,\tau_{y}\Bigr\},
\end{split}
\label{Ham-cont}
\end{align}
where $\{A,B\}=AB+BA$ denotes the anticommutator, $\tau_{0}$ is the $2\times2$ identity in Nambu space, and the separation of the $\sigma_{y}$ term from $\tau_{z}$ follows from $\sigma_{y}^{T}=-\sigma_{y}$ because the hole block is the negative transpose of the electron block. The phase profile has $\theta_L=0$ for $x<0$ and $\theta_R=\theta$ for $x>0$. On each side $\theta$ is constant, so $\{p_{x},f(\theta)\}=2f(\theta)p_{x}$ and the pairing reduces to the standard $p$-wave form. We therefore define the phase difference across the junction as
\begin{align}
\theta=\theta_R-\theta_L.
\end{align}
The profile $\delta\mu(x)=-\operatorname{sgn}(x)\delta\mu$ describes a static leg-dependent band-structure asymmetry across the junction. The superconductors remain at zero voltage bias, so $\theta$ is time independent and the Josephson current is an equilibrium dc current. The symmetry properties of this Hamiltonian are analyzed in Appendix~\ref{Symmetry Analysis}, indicating that $\phi$ and $\delta\mu$ play a key role in the diode effect.

% Table I removed: the conditions for a diode response are stated as text below.
A pronounced diode response is favored when current-reversing symmetries are absent and contributions with different gauge-invariant phases and CPR structures have comparable amplitudes.

\subsection*{Small interleg coupling: emergence of phase difference.}

To understand the origin of the diode effect, we focus on the regime of small $t_\perp$, where opposite-band tunneling processes become important. 
In this regime, it is convenient to express the junction coupling in terms of the weak-link Hamiltonian introduced in Eq.~\eqref{tunneling}. 
In the continuum limit, the tunneling term can be written in momentum space as
\begin{align}
H_{\Gamma}
=\Gamma\sum_{\alpha}\int_{-\infty}^{+\infty}dk dk'\left[\psi_{L,\alpha}^{\dagger}(k)\psi_{R,\alpha}(k')+h.c.\right].
\end{align}
Utilizing the unitary transformation in Eq.~\eqref{unitary transformation}, the tunneling term reads,
\begin{align}
\begin{split}
H_{\Gamma}=&\Gamma\sum_{\nu}\int_{-\infty}^{+\infty}dk dk'\left[\frac{t_\perp}{\sqrt{\delta \mu^2+t_{\perp}^2}}\psi_{L,\nu}^{\dagger}(k)\psi_{R,\nu}(k')\right.\\
&\left.+\nu e^{-\nu\frac{i\phi}{2}}\frac{\delta \mu}{\sqrt{\delta \mu^2+t_{\perp}^2}}\psi_{L,\nu}^{\dagger}(k)\psi_{R,-\nu}(k')+h.c.\right].  
\end{split} \label{gamma-coupling}
\end{align}
These renormalized tunneling terms separate the coupling into same-band and opposite-band contributions which, in the following, we will refer to as the same-band ($p_{\nu}/p_{\nu}$) and opposite-band ($p_{\nu}/p_{-\nu}$) tunneling sectors, respectively. Because $\delta\mu(x)=-\operatorname{sgn}(x)\delta\mu$, the two sides of the junction have mixing angles
\begin{align}
\kappa_L=\kappa,\qquad
\kappa_R=\pi-\kappa,
\end{align}
with $\kappa$ defined through $\cos\kappa=\delta\mu/r$, $\sin\kappa=t_\perp/r$. Each band carries an effective pairing phase $\phi_{\nu,\mathrm{eff}}(\phi,\kappa)$ defined in Eq.~\eqref{effective phase of the intraband pairing gap functio}.

For a tunneling amplitude $T_{ab}$ from band $a$ on the left to band $b$ on the right, the gauge-invariant Josephson channel phase is
\begin{align}
\Theta_{a/b}
=
\theta
-
\phi_{b,\mathrm{eff}}^R
+
\phi_{a,\mathrm{eff}}^L
-
2\arg T_{ab}
\quad(\mathrm{mod}\ 2\pi).
\label{gauge invariant channel phase}
\end{align}
The term $2\arg T_{ab}$ accounts for the Cooper-pair tunneling phase, which is twice that of the single-electron matrix element. The band operators also depend on the gauge used to diagonalize the leg Hamiltonian. Our convention defines $\phi_{\nu,\mathrm{eff}}$ through the upper-right BdG pairing block. The annihilation pairing coefficient $-\Delta_{k,\nu}^{*}$ therefore enters $\Theta_{a/b}$ with the opposite phase.

\paragraph{Same-band $p_\nu/p_\nu$ channel.}
From Eq.~\eqref{gamma-coupling}, $T_{\nu\nu}=\Gamma t_\perp/r$ is real, so $\arg T_{\nu\nu}=0$. The relevant phase factors on the left and right sides are
\begin{align}
\begin{split}
A_\nu^L
=&
\cos^2\frac{\kappa}{2}
+
\sin^2\frac{\kappa}{2}e^{-i\nu\phi},\\
A_\nu^R
=&
\sin^2\frac{\kappa}{2}
+
\cos^2\frac{\kappa}{2}e^{-i\nu\phi}.
\end{split}
\end{align}
A direct computation gives
\begin{align}
A_\nu^R A_\nu^{L*}
=
X(\phi,\kappa)
-i\nu Y(\phi,\kappa),
\end{align}
with
\begin{align}
\begin{split}
X(\phi,\kappa)
=&
\cos\phi
+
\sin^2\kappa\,
\sin^2\frac{\phi}{2},\\
Y(\phi,\kappa)
=&
\cos\kappa\sin\phi.
\end{split}
\end{align}
Using the same principal-value convention, the gauge-invariant same-band channel phase is defined when $X^2+Y^2>0$ and reads
\begin{align}
\Theta_{\nu/\nu}
=\theta+\nu\operatorname{Arg}(X+iY)
\quad(\mathrm{mod}\ 2\pi).
\label{effective phase difference}
\end{align}
Under the convention used here, $\delta\mu_L=+\delta\mu>0$, the weak-coupling limit corresponds to $\kappa_L\to0$ and $\kappa_R\to\pi$. Therefore,
\[
\Theta_{\nu/\nu}
\to
\theta+\nu\phi.
\]
The lattice low-energy band edge near $k=\pi$ has negative curvature. Mapping it to the positive-mass continuum description uses the corresponding hole representation, with $\mu_\alpha^{(h)}=2t_\alpha-\mu_\alpha$ and conjugated superconducting phases. For the lattice parameters below, this reverses the sign of the left-side detuning relative to the continuum convention, giving the weak-coupling phase combination $\theta+\phi$ and a crossing near $\theta\simeq\pi-\phi$. In the strong-coupling limit $t_\perp\gg|\delta\mu|$ ($\kappa\to\pi/2$), $\Theta_{\nu/\nu}\to\theta$.

\paragraph{Opposite-band $p_\nu/p_{-\nu}$ channel.}
From Eq.~\eqref{gamma-coupling}, $T_{\nu,-\nu}=\nu\Gamma(\delta\mu/r)e^{-i\nu\phi/2}$, so $2\arg T_{\nu,-\nu}=-\nu\phi\pmod{2\pi}$. The effective pairing phases satisfy the exact identity
\begin{align}
A_{-\nu}(\phi,\pi-\kappa)
=
e^{i\nu\phi}A_\nu(\phi,\kappa),
\end{align}
which implies $\phi_{-\nu,\mathrm{eff}}(\phi,\pi-\kappa)-\phi_{\nu,\mathrm{eff}}(\phi,\kappa)=\nu\phi\pmod{2\pi}$. The negative of this BdG pairing-phase difference in Eq.~\eqref{gauge invariant channel phase} exactly cancels the contribution $-2\arg T_{\nu,-\nu}$, yielding
\begin{align}
\Theta_{\nu/-\nu}=\theta
\quad(\mathrm{mod}\ 2\pi).
\end{align}
Thus, within the normal-band channel decomposition, the opposite-band channel carries no independent gauge-invariant phase shift. Under the conditions of Sec.~\ref{Josephson Junction}, a diode response can arise from the interference of the phase-shifted same-band contribution with the approximately unshifted opposite-band contribution.
\paragraph{Diode effect in the topological phase.}

First consider the topologically nontrivial phase, i.e., $\sqrt{\delta \mu^2+t_{\perp}^2}>|\mu|$.  The first and second terms in Eq.~\eqref{gamma-coupling} connect the same and opposite normal-state band labels, respectively. In the parameter regime considered below, their low-energy sectors can have different proximity to the topological transition. In the weak-mixing topological regime, the low-energy current is often dominated by a Majorana-derived contribution whose normal-band content is primarily associated with the $p_{-}/p_{-}$ same-band component. However, if $t_{\perp}\ll|\delta\mu|$, the first tunneling term is much smaller than the second because $p_{-}/p_{-}$ coupling is suppressed by the factor $t_\perp /\sqrt{\delta \mu^2+t_{\perp}^2}$. The Majorana-derived contribution to the current is proportional to 
\begin{align}
I_{-/-}(\theta)\propto \frac{t_\perp}{r}\frac{\partial}{\partial\theta}\left|\cos\!\left(\frac{\Theta_{-/-}}{2}\right)\right|
\end{align}
In the weak-coupling limit, $\Theta_{-/-}\simeq\theta-\phi$, resulting in a gauge-invariant phase shift of the current contributed by the $p_{-}/p_{-}$ coupling. The $p_{+}/p_{+}$ same-band contribution is negligible in the parameter regime considered here. The $p_{\nu}/p_{-\nu}$ currents carry the unshifted phase $\Theta_{\nu/-\nu}=\theta$ and are approximately proportional to $\sin{\theta}$. The total current can then be approximately expressed as
\begin{align}
I(\theta)\propto \frac{t_\perp}{r}\frac{\partial}{\partial\theta}\left|\cos\!\left(\frac{\Theta_{-/-}}{2}\right)\right|+C\sin{\theta},
\end{align}
where $C$ is a function of $\cos{\kappa}$. The non-sinusoidal, phase-shifted $p_{-}/p_{-}$ contribution and the approximately sinusoidal unshifted contribution can jointly generate a Josephson diode effect when their amplitudes are comparable.  As our discussion shows, the phase difference between the parallel chains is essential for the realization of the diode effect. If $\phi=0$ or $\pi$, the diode effect vanishes, which would correspond to the restoration of the effective spinless time-reversal-like symmetry.

\paragraph{Diode effect near the topological transition and in the decoupled two-channel regime.}

In the topologically trivial phase, no MZMs exist. In the parameter regime sufficiently far from the topological transition, the phase-shifted low-energy contribution is strongly suppressed, and the calculated diode response vanishes. The finite-$t_\perp$ class-D bulk transition is defined by the gap-closing condition
\begin{align}\label{toptrans}
\sqrt{\delta\mu^2+t_{\perp}^2}=|\mu|.
\end{align}
At the exactly decoupled point $t_\perp=0$, the two legs have independent class-D invariants $\mathcal M_1$ and $\mathcal M_2$. The interval $0<\delta\mu/|\mu|<1$ at $t_\perp=0$ corresponds to the regime where both continuum chains are individually topological (the ``decoupled two-channel topological regime''). Here $\mathcal M_1=\mathcal M_2=-1$, so the total class-D parity $\mathcal M=\mathcal M_1\mathcal M_2=+1$ is trivial, yet two Majorana modes are present at each end, protected by the exact leg decoupling. When a small $t_\perp$ is turned on, these modes can hybridize and gap out without a class-D bulk gap closing, so this is a crossover rather than a distinct topological phase boundary.

If the system is close to the true class-D transition of Eq.~\eqref{toptrans} but far from the decoupled regime, the gap closes at $k=0$ in the $\lambda=-$ BdG branch, while the $\lambda=+$ branch remains gapped there. The label $\lambda$ here refers to the BdG excitation branches, not to the normal-band index $\nu$. In this case the current-phase relation is dominated by first-harmonic terms of the form $A\sin\theta+B\cos\theta$, which combine into a single shifted sinusoidal CPR $C\sin(\theta+\theta_0)$. Such a CPR may exhibit an anomalous phase offset $\theta_0$ but still satisfies $I(\theta)=-I(-\theta-2\theta_0)$, implying $|I_c^+|=|I_c^-|$ and therefore no diode effect. However, if the system is also near the decoupled two-channel regime (i.e., $t_\perp$ is small and both leg chemical potentials lie within their individual topological windows), then the $p_{+}$-derived sector can also approach low energies and cannot be neglected.

In~\cite{JosephsoneffectsbetweentheKitaevladdersuperconductors}, it was shown that if two legs are in the topologically nontrivial phase, with a small interleg hopping term $t_{\perp}$, the CPR is not simply $I(\theta)\propto\sin\theta$. In the decoupled limit, the Majorana-derived junction branches have a $\cos(\theta/2)$ phase dependence without an additional effective phase shift. A small interleg hopping splits these branches. To illustrate this splitting, we consider a schematic four-Majorana model with positive quasiparticle energies
\begin{align}\label{split energy0}
\varepsilon_{\pm}(\theta)=\left|\epsilon_{\perp}\pm J_M\cos\frac{\theta}{2}\right|.
\end{align}
Here $\epsilon_{\perp}\geq0$ and $J_M\geq0$ are effective Majorana coupling amplitudes obtained by projecting the bare interleg hopping $t_{\perp}$ and junction tunneling $\Gamma$ onto the low-energy modes, respectively. Equation~\eqref{split energy0} describes the level splitting in the projected four-Majorana model, with couplings set by the low-energy wave functions.
Within this schematic model, the zero-temperature contribution from these two levels is
\begin{align}
F_{\rm low}(\theta)=-\frac12\bigl[\varepsilon_{+}(\theta)+\varepsilon_{-}(\theta)\bigr].
\end{align}
For $\epsilon_{\perp}>J_M$, the absolute values are resolved uniformly: $\varepsilon_{\pm}=\epsilon_{\perp}\pm J_M\cos(\theta/2)$. Then $F_{\rm low}(\theta)=-\epsilon_{\perp}$ is constant, and the corresponding low-energy current $I_{\rm low}(\theta)=(2e/\hbar)\,\partial F_{\rm low}/\partial\theta$ vanishes within this model. For $\epsilon_{\perp}<J_M$, the arguments inside the absolute values change sign at $\cos(\theta/2)=\pm\epsilon_{\perp}/J_M$, producing a piecewise phase dependence and a finite Josephson current from this low-energy sector. These conditions concern the effective couplings in the schematic model, not the topological phase boundary of the full ladder.

\begin{figure}[!htbp]
\centering
\includegraphics[width=85mm]
{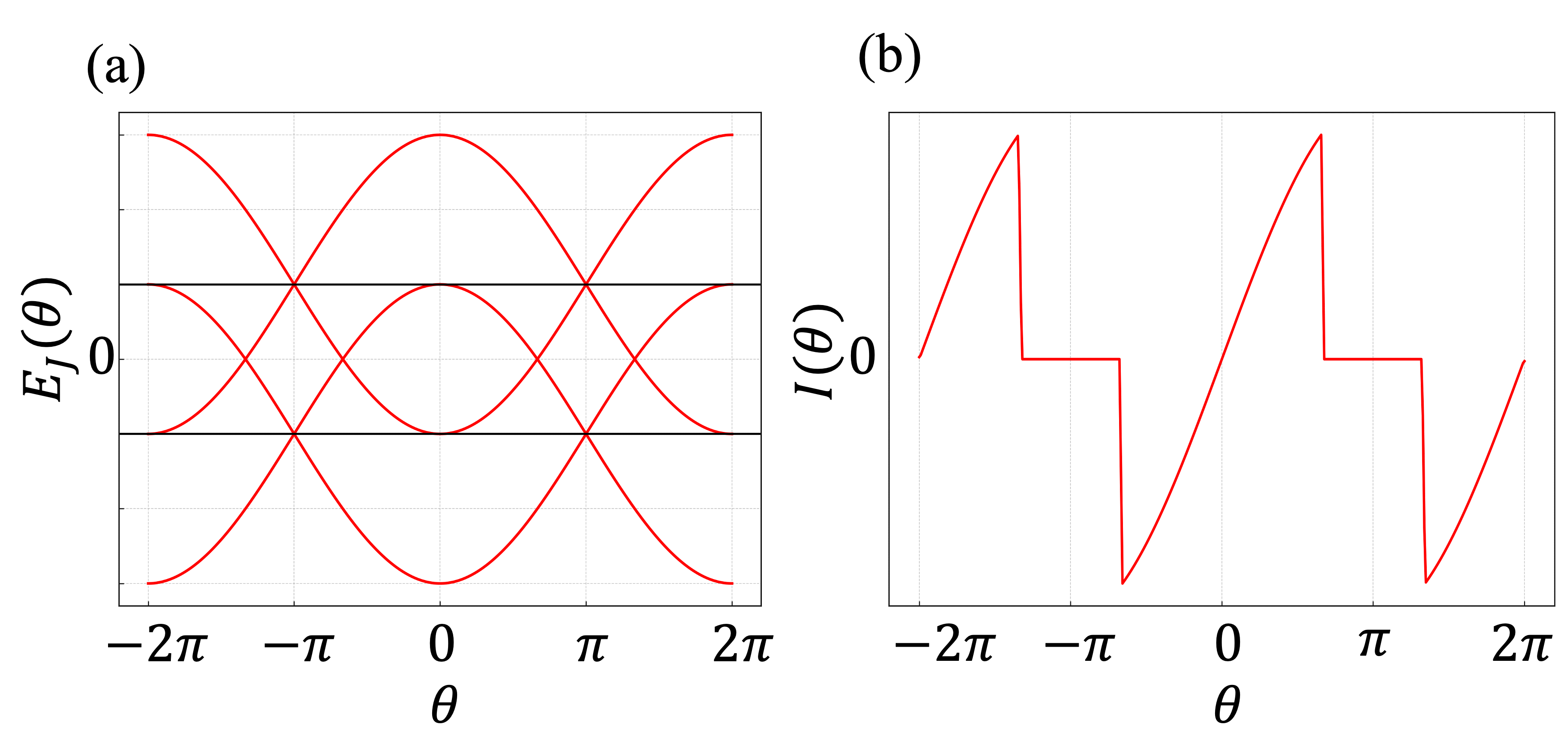}
\caption{Schematic four-Majorana model: (a) the signed particle-hole-symmetric energy branches $E=\pm[\epsilon_{\perp}\pm J_M\cos(\theta/2)]$ and (b) the corresponding low-energy CPR for $J_M=2\epsilon_{\perp}$. The independent nonnegative quasiparticle energies are $\varepsilon_{\pm}$ of Eq.~\eqref{split energy0}. The horizontal lines in (a) mark $E=\pm\epsilon_{\perp}$.} \label{transition energy phase}
\end{figure}

Near the decoupled two-channel regime, for sufficiently small $t_\perp$ satisfying $\sqrt{\delta\mu^2+t_\perp^2}<|\mu|$, the ladder remains on the class-D-trivial side, and no protected Majorana zero modes are present. Nevertheless, hybridized low-energy states continuously connected to the decoupled-limit Majorana sector can remain near zero energy and contribute to the Josephson current. Components originating from the opposite-band tunneling term contribute to these hybridized states, but this does not imply an independent gauge-invariant phase shift of a separate $p_\nu/p_{-\nu}$ channel.

As discussed in Eq.~\eqref{effective phase difference}, an effective phase difference occurs in the same-band channels on both sides of the topological transition. Consequently, the same-band-derived and opposite-band-derived spectral components can jointly affect the diode response. Our numerical results show that the same-band-derived component exhibits a nonmonotonic dependence on $t_{\perp}$ near the class-D topological transition described by Eq.~\eqref{toptrans} and the decoupled two-channel regime. In agreement with the mechanism of Sec.~\ref{Josephson Junction}, the diode response is largest where these contributions interfere under the stated conditions.

\subsection*{Large interleg coupling: suppression of the relative phase and approximate recovery of current-reversal constraints.}

For completeness, we demonstrate that the diode effect disappears if $t_{\perp}$ is large. From the symmetry perspective, the effect of the left-right asymmetry becomes perturbatively weak in the band basis (see Appendix~\ref{Symmetry Analysis} for details). After the Bogoliubov transformation, the point-contact tunneling Hamiltonian contains normal and anomalous terms with coherence factors depending on the momenta on both sides of the junction. The labels $p_\nu/p_\nu$ and $p_\nu/p_{-\nu}$ remain useful for identifying the normal-band components from which these terms originate. An individual BdG eigenstate generally contains several such components, however, so this organization is not an exact decomposition into independent physical currents. The explicit transformed Hamiltonian is not required for the phase argument: Eqs.~\eqref{effective phase difference} and the opposite-band result above establish that the same-band phase shift vanishes in the strong-coupling limit, whereas the opposite-band tunneling term carries no independent gauge-invariant phase shift.

In the topologically nontrivial regime considered here, the low-energy current is dominated by a Majorana-derived sector whose normal-band content is primarily associated with the $p_{-}/p_{-}$ same-band component. The lower normal-state band, with dispersion $\xi_{k,-}=\hbar^2 k^2/(2m)-\mu-r$, crosses the Fermi level at
\begin{align}\label{fermi-momentum-minus-band}
k_{F,-}=\frac{\sqrt{2m(\mu+r)}}{\hbar},
\end{align}
provided $\mu+r>0$. For the quasi-classical approximation~\cite{Geometry}, we assume that the band Fermi energy $E_{F,-}=\mu+r$ is large compared with the intraband pairing gap at the Fermi momentum, $E_{F,-}\gg|\Delta_{k_{F,-},-}|$. The corresponding low-energy Josephson energy is
\begin{align}
E(\theta)= - \frac{|\Delta_{k_{F,-},-}|}{\sqrt{1+Z^2}} \left|\cos\frac{\theta}{2}\right|,
\end{align}
where $Z=mU_{0}/(\hbar^2 k_{F,-})$. This yields the reciprocal CPR $I(\theta)\propto\operatorname{sgn}[\cos(\theta/2)]\sin(\theta/2)$.
In both the topologically trivial and nontrivial cases, the largest Josephson currents are the same in both current directions, because $ I(\theta) = - I (-\theta)$.

Class-D triviality alone does not determine whether the normal-state bands cross the Fermi level. In the large-$t_\perp$ regime considered here, the same-band phase shift vanishes and no independent interband phase shift remains. The remaining weak-link contributions are therefore conventional gapped Andreev processes whose lowest-order CPR is sinusoidal, $I(\theta)\propto\sin\theta$. Consequently, the CPR approaches a current-reversal-symmetric form and the diode response is suppressed, vanishing in the strong-coupling limit.
%Thus, in either topological trivial or nontrivial case, %diode effect vanishes with a large $t_{\perp}$. 

\section{Numerical Results}\label{Numerical Results}
In this section we present the numerical results from exact diagonalization of the BdG Hamiltonian to substantiate our analytical findings and also cover a wider parameter range.

\begin{figure}[H]
\centering
\includegraphics[width=0.98\columnwidth]
{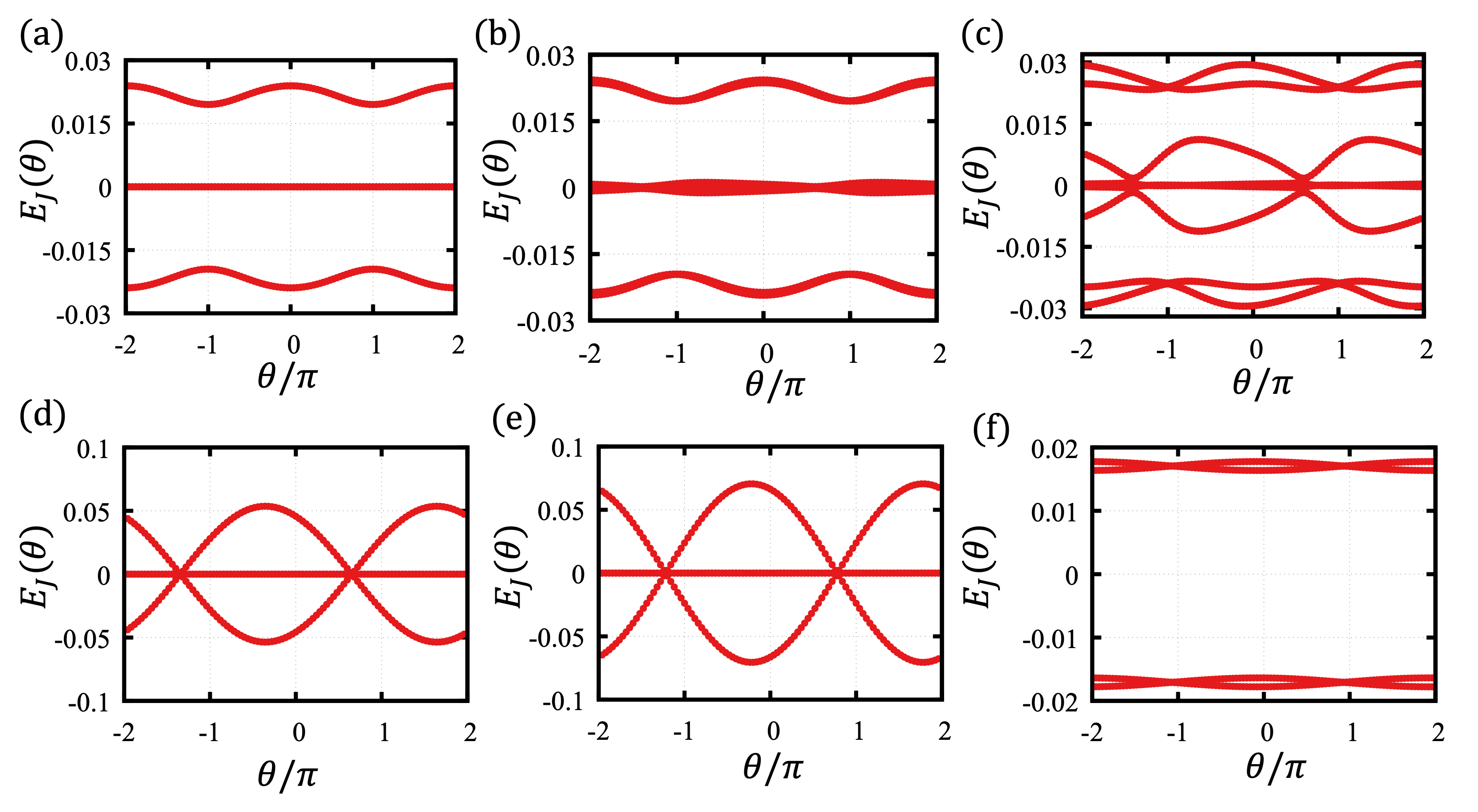}
\caption{The low-energy BdG spectrum near $E=0$ as a function of $\theta$. Parameters: $\mu_{1}=0$, $t_{1}=\mu_{2}=2t_{2}=1$, $\Delta=0.5$, $\phi=\pi/3$, $\Gamma=0.1$ (in units of $t_1$). (a)-(f) $t_{\perp}=0$, 0.01, 0.1, 0.5, 1.0, 2.0.} \label{fig:5}
\end{figure}

\begin{figure}[!htbp]
\centering
\includegraphics[width=0.98\columnwidth]
{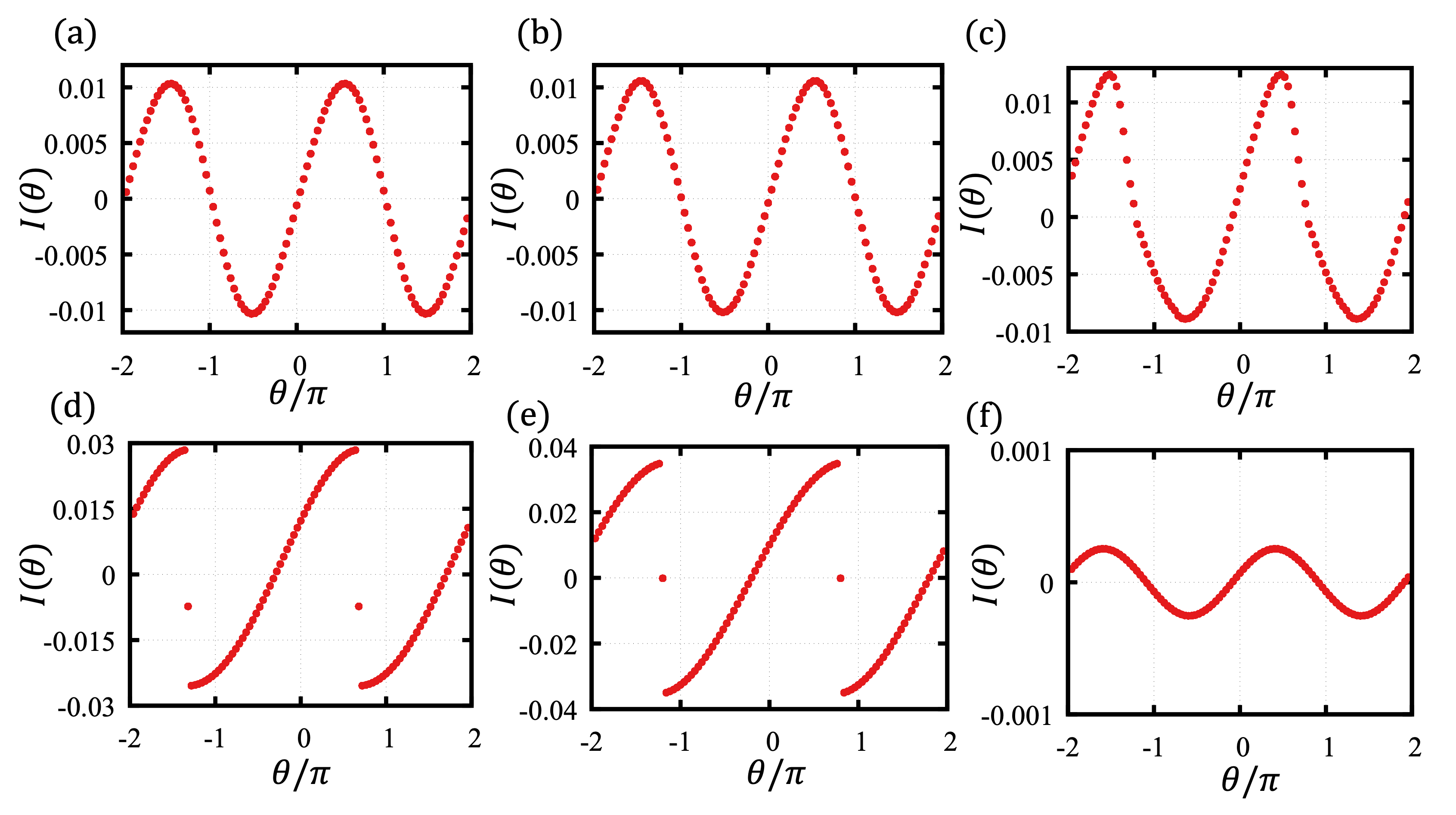}
\caption{The total Josephson CPR. Parameters as in Fig.~\ref{fig:5}. (a)-(f) $t_{\perp}=0$, 0.01, 0.1, 0.5, 1.0, 2.0.} \label{fig:3}
\end{figure}

\begin{figure}[H]
\centering
\includegraphics[width=0.98\columnwidth]
{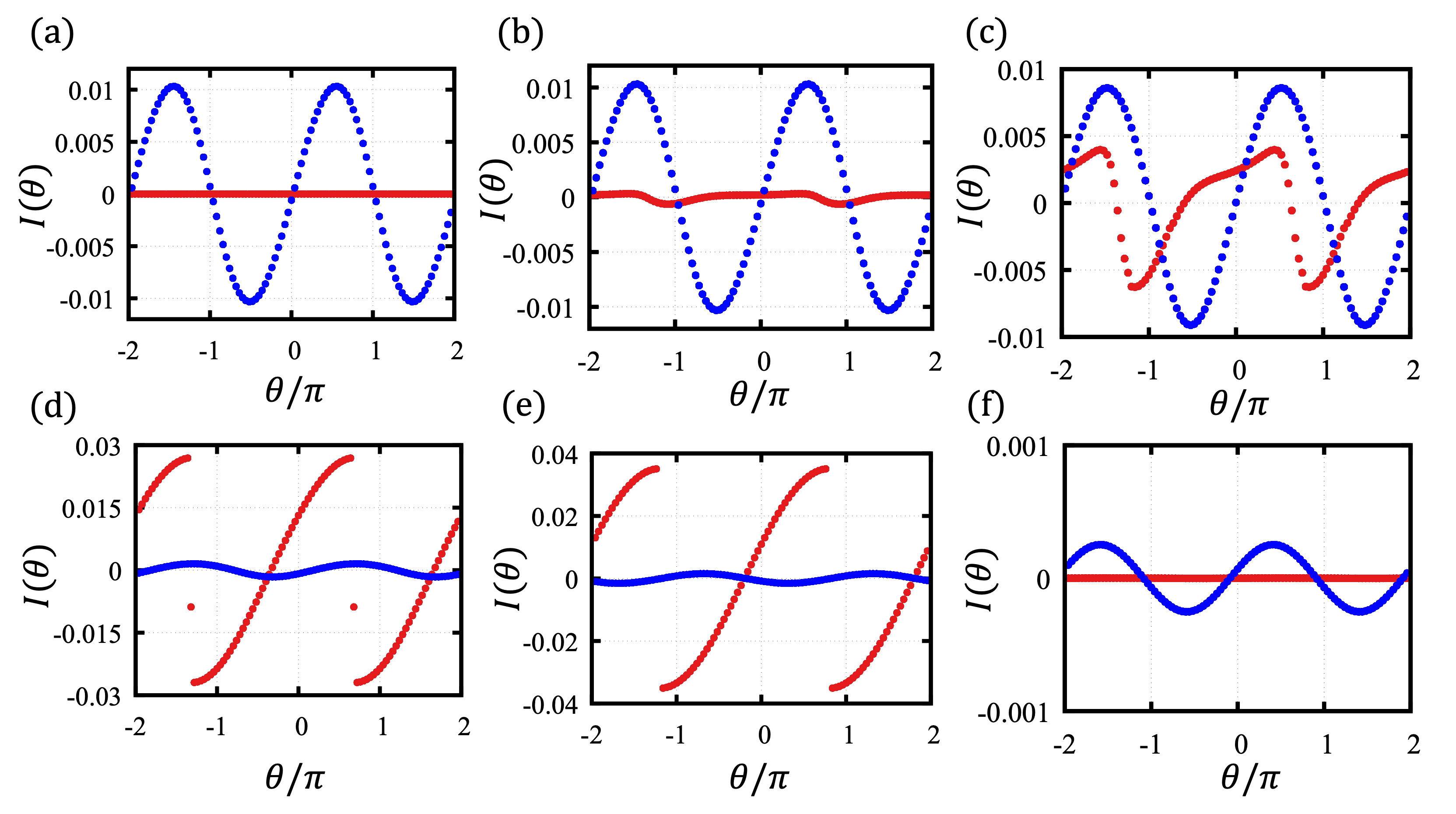}
\caption{The decomposition of the Josephson CPR for the parameters and panel order of Fig.~\ref{fig:5}. Red: the contribution of the two lowest positive-energy BdG levels. Blue: the remaining spectral contribution.} \label{fig:4}
\end{figure}

\begin{figure}[!htbp]
\centering
\includegraphics[width=0.98\columnwidth]
{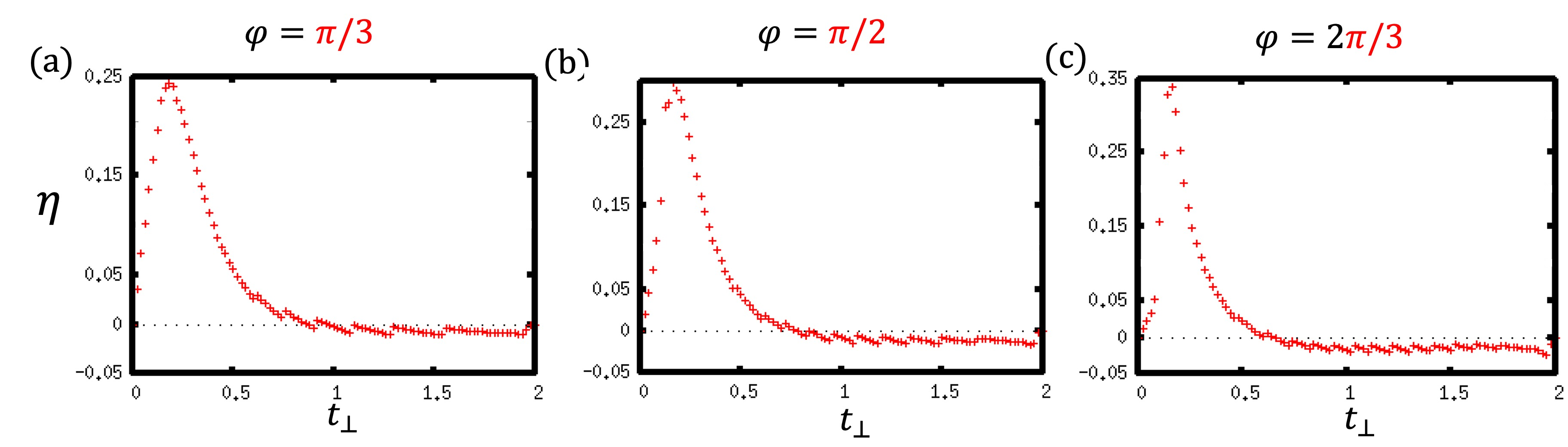}
\caption{The efficiency of the Josephson diode effect as a function of $t_{\perp}$. Parameters as in Fig.~\ref{fig:5}. (a)-(c) $\phi=\pi/3$, $\pi/2$, $2\pi/3$.} \label{fig:diodeefficiency}
\end{figure}

The numerical calculations are performed on a lattice with 100 rungs on each side of the junction (each rung containing two sites, one on each leg), with open boundary conditions at the outer ends. The BdG Hamiltonian matrix dimension is therefore $800\times800$ (400 physical fermionic sites, Nambu-doubled). The energy unit is set by $t_1\equiv1$. All parameters are quoted in these units: $t_1=1.0$, $t_2=0.5$, $\mu_1=0$, $\mu_2=1.0$, $\Delta=0.5$, $\phi=\pi/3$, $\Gamma=0.1$, with $t_\perp$ varied for Figs.~\ref{fig:5}--\ref{fig:diodeefficiency}. At zero temperature, the Josephson ground-state energy is
\begin{align}\label{Fdef}
F(\theta)=-\frac12\sum_{E_n(\theta)>0}E_n(\theta),
\end{align}
where $E_n(\theta)$ are the positive BdG eigenvalues. The equilibrium current is $I(\theta)=(2e/\hbar)\,\partial_\theta F(\theta)$, evaluated numerically and expressed in units of $2e t_1/\hbar$. To examine the low-energy contribution to Fig.~\ref{fig:4}, we use the two lowest positive-energy levels $E_1(\theta)\le E_2(\theta)$:
\begin{align}
F_{\mathrm{low}}(\theta)&=-\tfrac12[E_1(\theta)+E_2(\theta)],\\
F_{\mathrm{rest}}(\theta)&=F(\theta)-F_{\mathrm{low}}(\theta).
\end{align}
The corresponding currents satisfy $I=I_{\mathrm{low}}+I_{\mathrm{rest}}$. This spectral decomposition characterizes the low-energy contribution without assigning each BdG state to a separate tunneling channel.

For the lattice calculation, we choose $\mu_1-2t_1<0$ and $\mu_2-2t_2=0$, placing one leg in the topological regime and the other at its transition. Their assignments are exchanged across the junction, as shown in Fig.~\ref{fig:ladder}. With these parameters, Eq.~\eqref{topological constraints0} gives $\mathcal M=-1$ for $0<t_\perp<2$ and $\mathcal M=+1$ for $t_\perp>2$, with a bulk transition at $t_\perp=2$. The diode response emerges at finite interleg coupling, reaches its maximum at intermediate coupling, and is suppressed at large $t_\perp$.
Fig. \ref{fig:5} shows the low-energy spectrum for varying $\theta $ obtained numerically for representative values of $ t_{\perp} $. Panel (a) depicts the case of decoupled Kitaev chains. This consists of two copies of a Josephson junction coupling a topologically nontrivial system with one at a topological transition state (a junction connecting a topological leg to a leg tuned to its bulk transition). With this setup of different topological phases, the inversion symmetry is broken. At $t_\perp=0$, the two legs are dynamically decoupled, allowing independent gauge choices for their superconducting phases. The effective complex-conjugation constraint is preserved, and the CPR remains reciprocal. The low-energy crossing is associated with the critical leg and the particular finite-junction spectrum: the Josephson energy $E_{J}(\theta)<0$ is proportional to $-\cos{\theta}$ with degenerate branches, indicating the ABSs have the same origin as those between a conventional superconductor and a topological superconductor at its bulk transition.

For the chosen parameters, the coupled ladder supports protected Majorana end modes for $0<t_\perp<2$. At $t_\perp=0.01$ [Fig.~\ref{fig:5}(b)], the spectrum contains a Majorana-derived branch crossing zero energy and a second, more conventional ABS branch. Interleg coupling hybridizes the normal-band components of these states. At $t_\perp=0.1$ [Fig.~\ref{fig:5}(c)], the low-energy phase dispersion increases and develops a shift consistent with the same-band tunneling mechanism of Sec.~\ref{Josephson Junction}. In particular, at $\phi=\pi/3$ the crossing near $\theta\approx2\pi/3$ agrees with the lattice-band-edge phase mapping discussed after Eq.~\eqref{effective phase difference}. The Majorana-derived low-energy branch remains prominent in panels (d) and (e). With stronger interleg coupling the diode response decreases, and for $t_\perp>2$ the chosen lattice ladder is class-D trivial.
Fig. \ref{fig:3} summarizes the CPR derived from the ground-state energy $F(\theta)$.

Figure~\ref{fig:4} shows how the low-energy and remaining spectral contributions combine in the equilibrium CPR. At $t_\perp=0$, the low-energy contribution vanishes and the current is reciprocal. Finite interleg coupling produces a shifted, non-sinusoidal low-energy response alongside a distinct contribution from the remaining states. Their competition produces unequal critical currents when the two contributions have comparable amplitudes. The sharp changes in panels (d) and (e) are consistent with the near-zero-energy crossings in Figs.~\ref{fig:5}(d) and (e). Both the low-energy contribution and the diode effect are suppressed at large $t_\perp$.

Next we consider the impact of the phase difference $ \phi $ between the legs, as shown in Fig. \ref{fig:diodeefficiency} for $\phi$ from $\pi/3$ to $2\pi/3$. Here we plot the efficiency of the diode effect,  
\begin{align} 
\eta=\frac{I_c^+-I_c^-}{I_c^++I_c^-}
\label{eq:efficiency}
\end{align}
as a function of $t_{\perp}$. Here $I_c^+$ ($I_c^-$) denotes the absolute value of the critical Josephson current in the positive (negative) direction, defined as $I_c^+=\max_\theta I(\theta)$ and $I_c^-=-\min_\theta I(\theta)>0$. For $t_\perp=0$ the efficiency is exactly zero.  We observe that the efficiency of the diode effect reaches a maximum at a small yet nonzero $t_{\perp}$ for which the amplitudes of the different current contributions are comparable with each other. We also see the impact of the phase difference $\phi$ on $ \eta $. For $\phi=2\pi/3$, $ \eta$ reaches about 35\%, the highest among the three cases displayed. 

The suppression of $\eta$ at large $t_\perp$ also follows from the symmetry analysis. The spatial profile $\delta\mu(x)$ breaks inversion $\mathcal P$ and the combined current-reversing constraints $\widetilde{\mathcal T}\mathcal M'_{x/y}$. As $t_\perp$ becomes large compared with $|\delta\mu|$, this asymmetry becomes perturbatively weak in the band basis and the CPR approaches a current-reversal-symmetric form. The diode efficiency is therefore largest where the symmetry breaking remains effective and the current contributions have comparable amplitudes.

We next focus on the parameter range around the topological transition, while keeping $ t_{\perp} >0 $ fixed. The aim is to emphasize the role the topological nature of the system plays.
 \begin{figure}[!htbp]
\centering
\includegraphics[width=40mm]
{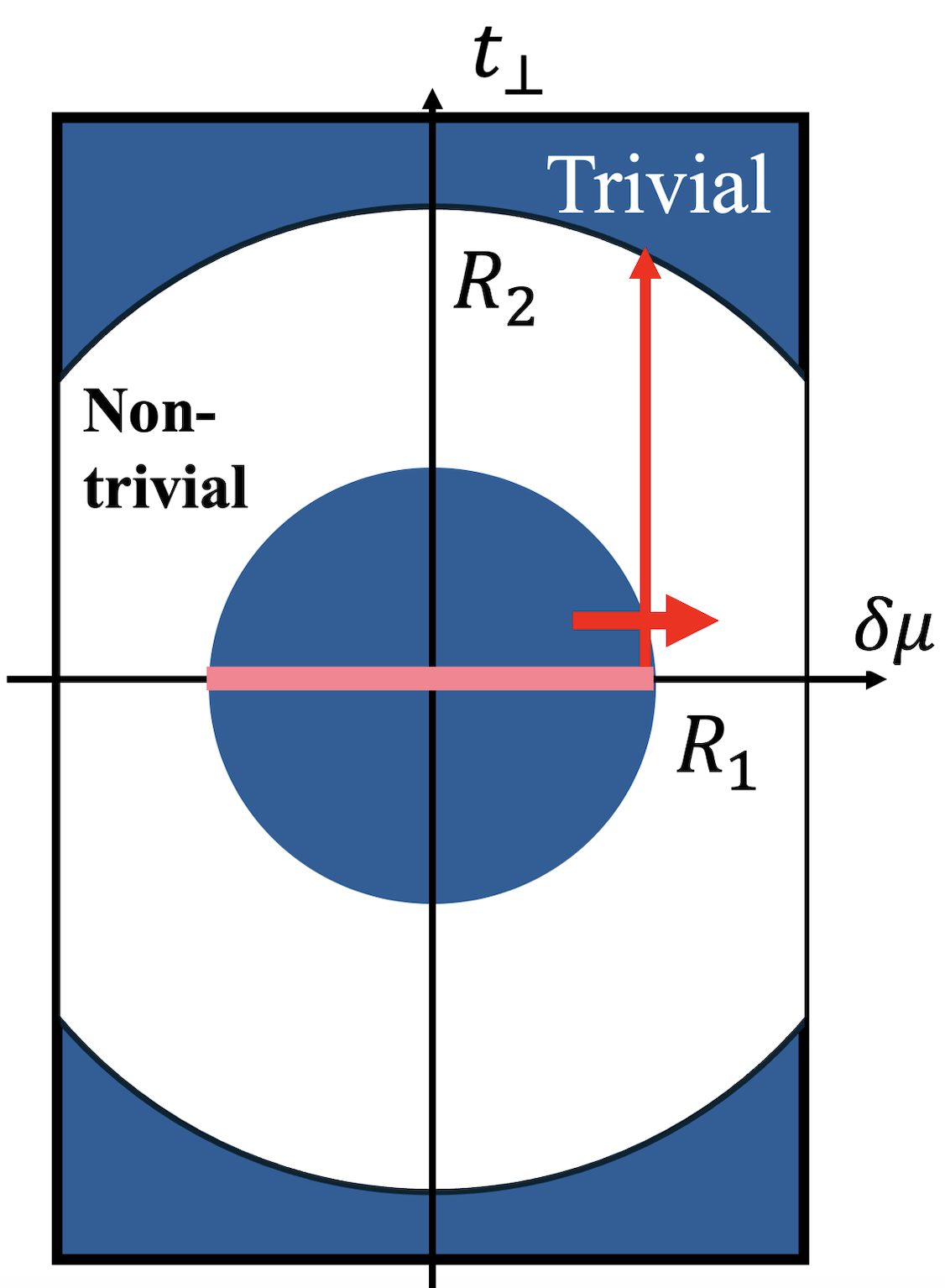}
\caption{The phase diagram of a Kitaev ladder as a function of the chemical potential difference $\mu_2-\mu_1$ and perpendicular hopping term $t_{\perp}$. The arrows indicate the changed parameters.} \label{fig:topologicalphasewitharrow}
\end{figure}
Figure~\ref{fig:topologicalphasewitharrow} shows the topological phase diagram of the Kitaev ladder.
To study the diode effect near the topological transition, we fix all parameters except $\mu_2$. For this scan we use a different parameter set: $t_1=t_2=1.0$, $\mu_1=0$, and $t_{\perp}=0.1$ (in units of $t_1$). The chemical potential $\mu_2$ is our tuning parameter, varied between 1.9 and 2.0. For vanishing $t_\perp$, both legs are independently topologically nontrivial in this range of $\mu_2$ (except for the endpoint $\mu_2=2.0$, which lies at the decoupled-chain transition). A small yet non-vanishing $t_\perp$, as examined in Appendix~\ref{Topological criterion}, moves the class-D topological transition into this $\mu_2$ range. For the choice $t_{\perp}=0.1$ the phase boundary occurs at $\mu_2^*=1.995$. Only for $\mu_2>\mu_2^*$ is the system topologically nontrivial, which allows us to explore the region around the transition.
\begin{figure}[H]
\centering
\includegraphics[width=0.98\columnwidth]
{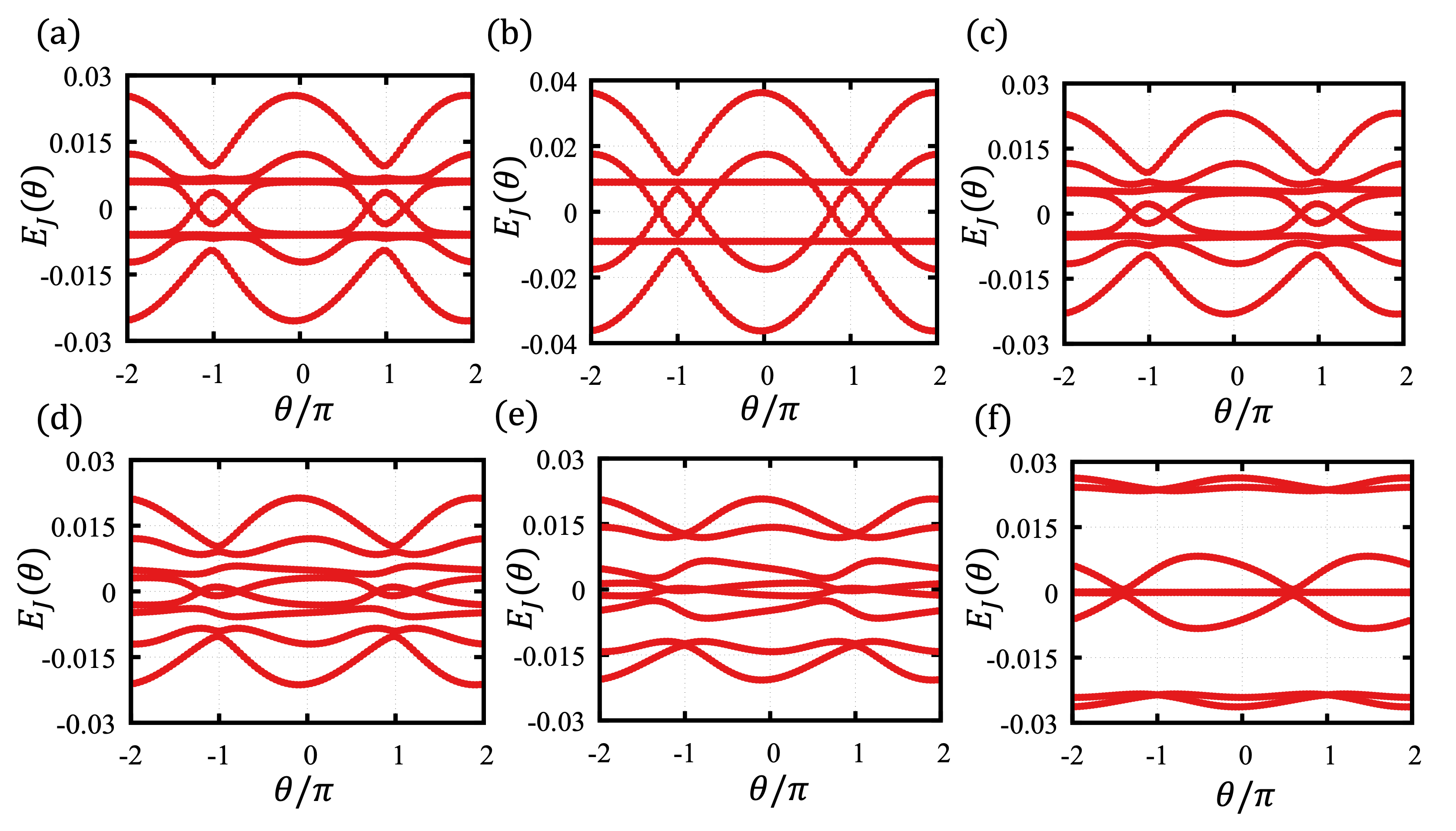}
\caption{The low-energy BdG spectrum near $E=0$ as a function of $\theta$. Parameters: $\mu_1=0$, $t_1=t_2=1.0$, $t_{\perp}=0.1$, $\Delta=0.5$, $\phi=\pi/3$, $\Gamma=0.1$ (in units where $t_1=1$). (a)-(f) $\mu_{2}=1.90$, 1.95, 1.96, 1.97, 1.98, 2.0.} \label{diodeenergytransition}
\end{figure}

In Fig.~\ref{diodeenergytransition}, we plot the low-energy BdG spectrum for different values of $\mu_2$ while fixing all other parameters. The panels (a)-(e) cover the system in the topologically trivial state, while for panel (f) the system is topologically nontrivial. In Fig.~\ref{diodeenergytransition}(a) ($\mu_2=1.9$), low-energy phase-dispersive branches are visible. These branches are inherited from the decoupled-junction limit but are split and hybridized by the non-vanishing interleg hopping, as illustrated schematically by Eq.~\eqref{split energy0} and Fig.~\ref{transition energy phase}(a). At $\mu_2=1.9$, the dominant low-energy branches show no appreciable effective phase shift, indicating that the phase-shifted same-band contribution remains weak in this regime. As the system approaches the topological boundary, the same-band-derived low-energy component develops a stronger phase dependence near $E=0$. With increasing $\mu_2$, the system approaches the class-D topological phase boundary determined by the lattice Pfaffian criterion in Eq.~\eqref{topological constraints0}. Near this transition, the same-band-dominated low-energy branch is strongly modified around $E=0$. The phase dependence of the low-energy branches changes as the system crosses the topological transition into panel (f).

\begin{figure}[!htbp]
\centering
\includegraphics[width=0.98\columnwidth]
{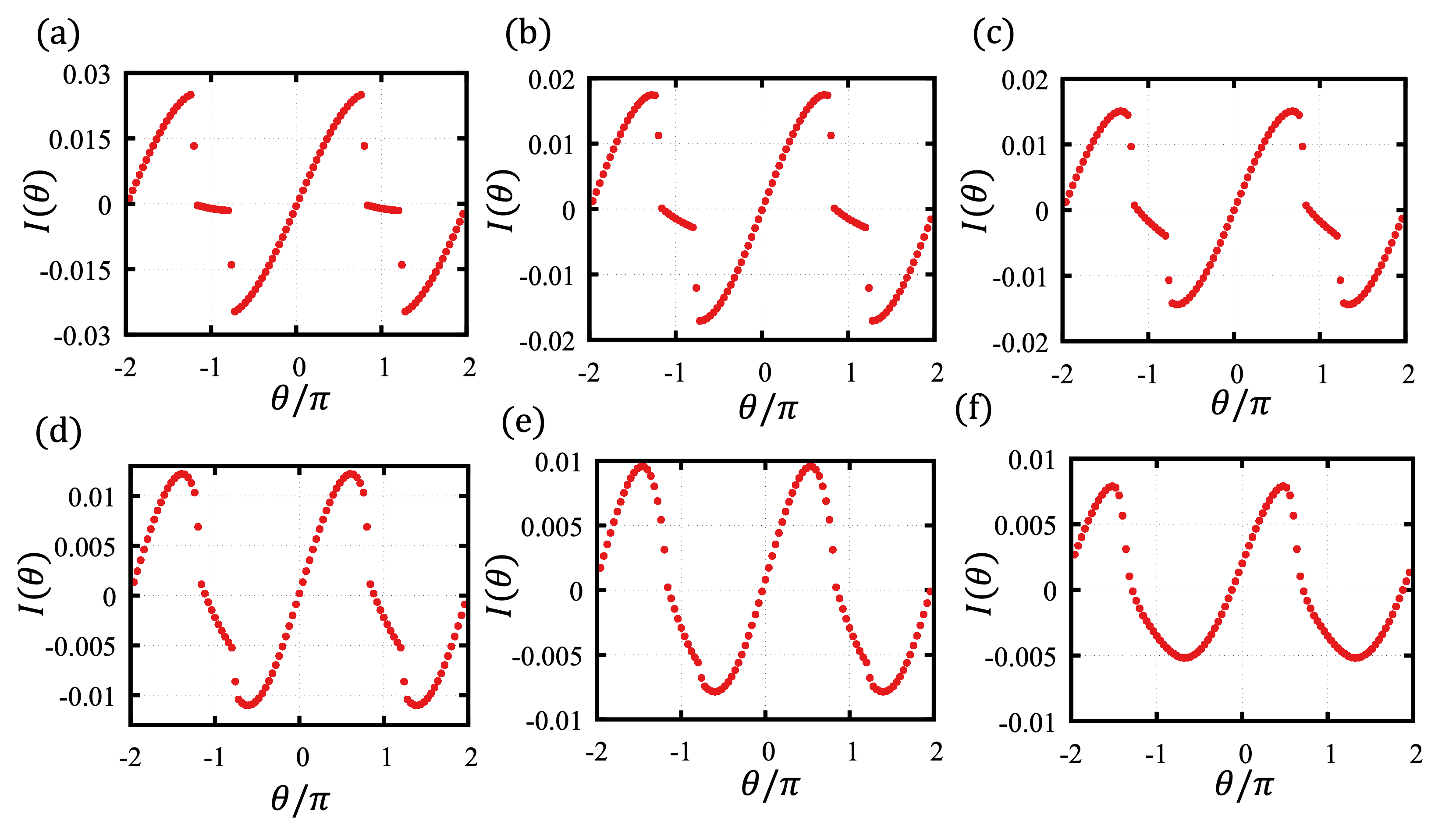}
\caption{The Josephson CPR corresponding to the spectra in Fig.~\ref{diodeenergytransition}. Parameters as in Fig.~\ref{diodeenergytransition}. (a)-(f) $\mu_{2}=1.90$, 1.95, 1.96, 1.97, 1.98, 2.0.} \label{diocurrenttransition}
\end{figure}

In Fig.~\ref{diocurrenttransition}, we plot the corresponding Josephson CPRs. The diode effect is essentially absent for $\mu_{2}=1.9$ because the phase-shifted same-band-derived low-energy component makes only a small contribution to the low-energy Josephson current. As $\mu_{2}$ approaches the transition, this component is enhanced through hybridization with low-energy states connected to the decoupled-chain Majorana sector. Its competition with the approximately unshifted ABS contribution then produces a finite diode response. Note that since an opposite-band-dominated low-energy branch approaches and crosses $E=0$ near the topological transition, the same-band-dominated current contribution is strongly enhanced through hybridization with the low-energy states inherited from the decoupled-chain Majorana sector.

\begin{figure}[!htbp]
\centering
\includegraphics[width=80mm]
{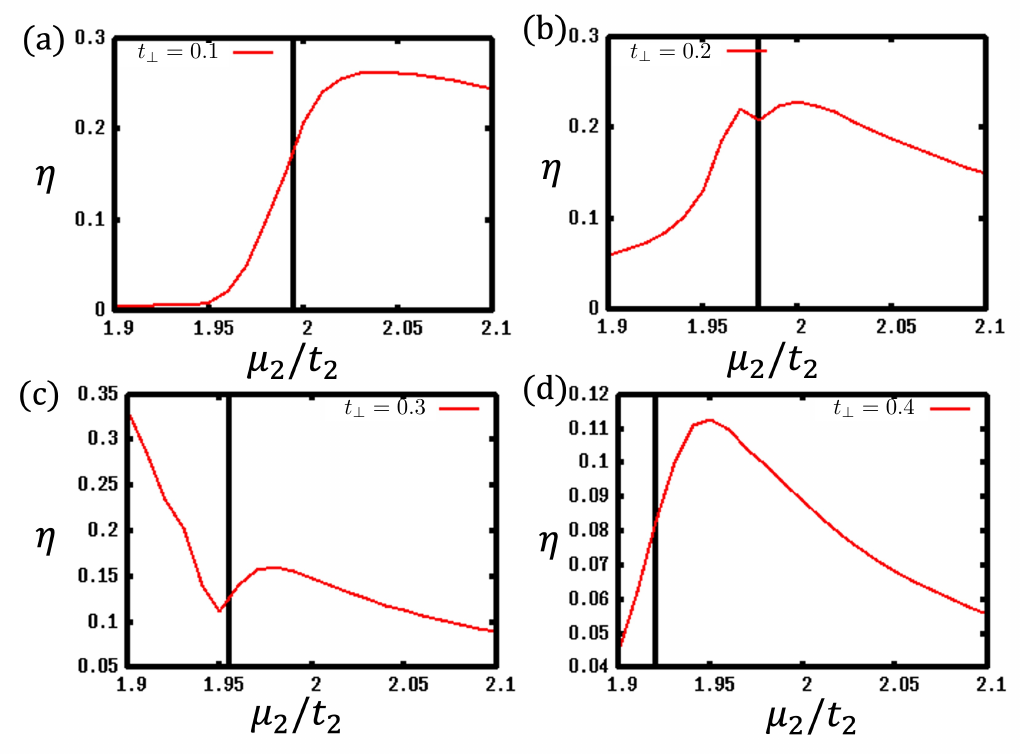}
\caption{The efficiency of the diode effect as a function of $\mu_2$. Parameters as in Fig.~\ref{diodeenergytransition} except $t_{\perp}$ varied. (a)-(d) $t_{\perp}=0.1$, 0.2, 0.3, 0.4. The vertical line indicates the position of the topological phase boundary.} \label{diodeefficiencytransition}
\end{figure}

In Fig.~\ref{diodeefficiencytransition} we display the behavior of the efficiency $ \eta $ defined in Eq.~(\ref{eq:efficiency})
as a function of $\mu_2$ for different values of $ t_{\perp} $. In all cases, we observe that in the topologically nontrivial region $\eta$ increases initially and falls off after passing through a maximum. As illustrated by the continuum model in Eq.~(\ref{gamma-coupling}),  in this context it is important to note that near the transition point a Majorana-derived low-energy contribution with a substantial phase-shifted same-band component becomes prominent and carries the same-band gauge-invariant phase shift. This contribution decreases away from the transition point.
 
On the topologically trivial side, the behavior is more complicated. While $\eta$ increases monotonically when approaching the transition point for case (a) and (d), we observe a dip for (b) and (c).
To understand the difference in this behavior, it is important to elucidate the interplay between current contributions with predominantly opposite-band- and same-band-derived normal-state content. In the calculated parameter range, the diode efficiency becomes largest when the relevant phase-shifted and approximately unshifted current contributions have comparable amplitudes. To understand why the efficiency increases in Fig.~\ref{diodeefficiencytransition}(a), we first note that for $t_{\perp} =0 $ the system consists of two isolated Josephson junctions with degenerate energies. These are split by finite $ t_{\perp} $ (see Fig.~\ref{diodeenergytransition}) without effective phase difference, and the same-band-derived component carries a distinct phase dependence as $\mu_2$ increases. This leads to the growing  diode efficiency near the topological transition boundary.

In Fig.~\ref{diodeefficiencytransition}(b) and (c), the efficiency instead decreases near the phase boundary. For larger $t_\perp$, the BdG eigenstates contain stronger mixtures of same-band- and opposite-band-derived components. Their phase-dependent current can therefore no longer be assigned to a single independent $p_{\nu}/p_{-\nu}$ channel with its own gauge-invariant phase shift. The complex CPR in this regime reflects hybridization of several contributions rather than a new irreducible opposite-band phase shift. Near the transition point, as the energy levels crossing zero disappear, this hybridization-induced contribution diminishes. As a consequence, the overall diode efficiency decreases in this regime.

If one continues to increase $t_{\perp}$, the opposite-band-dominated zero-energy crossing is almost completely removed. The remaining hybridization-induced phase-dependent contribution alone should not give rise to a diode effect once the complementary contribution with a different CPR structure is suppressed. The distinct gauge-invariant phase dependence is then dominated by the same-band Majorana-derived component.

We note that the present analysis is carried out in the clean equilibrium limit. The influence of disorder on the diode response, which may in particular generate trivial low-energy states in realistic finite-size implementations, remains an open problem beyond the scope of this work.

\section{Conclusion}\label{Conclusion}
We investigated the equilibrium Josephson diode effect in a junction between two Kitaev ladders. The interleg phase difference and the left--right leg asymmetry remove the relevant current-reversing constraints. In the normal-band description, same-band tunneling acquires a gauge-invariant phase shift, while opposite-band tunneling carries the bare junction phase. Their different CPR structures and comparable amplitudes can produce unequal positive and negative critical currents. The BdG calculations show that Majorana-connected low-energy states influence this interference in both the topological regime and near the class-D transition, alongside the contributions of the remaining Andreev spectrum. The diode efficiency reaches a maximum at intermediate $t_\perp$ and is suppressed toward both the decoupled and strong-coupling limits.

A possible coupled-wire implementation would use barrier gates to tune $t_\perp$, phase-biased superconducting contacts to control $\phi$, and local gates to set the leg chemical potentials. Simultaneous independent control of these parameters remains experimentally challenging.

\section{Acknowledgment}
This work is financially supported by Gp-spin Program. C.-R. X. is grateful for the hospitality of the Pauli Center and the Institute for Theoretical Physics during his stay at ETH Zurich. M.S. is grateful for the financial support by the Swiss National Science Foundation (SNSF) through Division II (No. 184739).

\section*{Data availability}
Numerical code and data are available at \url{https://github.com/Slabestone/kitaev-ladder-josephson-diode}.

\appendix
\section{Derivation of the Majorana number}
\label{Derivation of the Majorana number}
The class-D Pfaffian invariant is determined at the particle-hole-invariant momenta $k=0,\pi$, where odd-parity pairing vanishes. We introduce the momentum-space Majorana operators
\begin{align}
\begin{split}
\gamma_{\alpha,A}(k)&=c_{\alpha}(k)+c_{\alpha}^{\dagger}(-k),\\
\gamma_{\alpha,B}(k)&=i\left[-c_{\alpha}(k)+c_{\alpha}^{\dagger}(-k)\right],
\end{split}
\end{align}
which satisfy the relation $\gamma_{\alpha,A/B}^{\dagger}(k)=\gamma_{\alpha,A/B}(-k)$.
At $k=0,\pi$, the Majorana-basis Hamiltonian takes the form
\begin{align}
\left.H_{\mathrm{lad}}\right|_{k=0,\pi} = \frac{i}{4} \sum_{k=0,\pi} \boldsymbol{\gamma}(k)^T \widetilde{\mathcal H}_{\mathrm{lad}}(k)\boldsymbol{\gamma}(k)
\end{align}
with the Majorana spinor
\begin{align}
\boldsymbol{\gamma}(k) = \begin{bmatrix}
\gamma_{1,A}(k)\\
\gamma_{1,B}(k)\\
\gamma_{2,A}(k)\\
\gamma_{2,B}(k)
\end{bmatrix}
\end{align}
and
\begin{align}
\widetilde{\mathcal H}_{\mathrm{lad}}(k)=
\begin{bmatrix}
0 & -\epsilon_{1,k} &0 &-t_{\perp} \\
\epsilon_{1,k} & 0  &t_{\perp} & 0\\
0 &-t_{\perp} &0 &-\epsilon_{2,k} \\
t_{\perp} & 0 &\epsilon_{2,k} & 0
\end{bmatrix}
\end{align}
where $\epsilon_{\alpha,k}=2t_\alpha\cos k+\mu_\alpha$. At $k=0,\pi$, the pairing terms vanish. The displayed skew-symmetric matrix is used only at these momenta and does not represent the general-$k$ Majorana kernel.

The Pfaffian of the $4\times4$ skew-symmetric matrix $\widetilde{\mathcal H}_{\mathrm{lad}}(k)$ is
\begin{align}
\operatorname{Pf}\widetilde{\mathcal H}_{\mathrm{lad}}(k)=\epsilon_{1,k}\epsilon_{2,k}-t_{\perp}^2.
\end{align}
At $k=0$ and $k=\pi$, the pairing term vanishes ($\Delta_{\alpha,0}=\Delta_{\alpha,\pi}=0$), so $\epsilon_{\alpha,0}=2t_{\alpha}+\mu_{\alpha}$ and $\epsilon_{\alpha,\pi}=-2t_{\alpha}+\mu_{\alpha}$.
The class-D $\mathbb{Z}_2$ topological invariant is the Majorana number $\mathcal{M}$, determined by the product of the signs of the Pfaffian at $k=0$ and $k=\pi$:
\begin{align}
\begin{split}
\mathcal{M}&=\operatorname{sgn}\!\bigl[\operatorname{Pf}\widetilde{\mathcal H}_{\mathrm{lad}}(0)\bigr]\;
\operatorname{sgn}\!\bigl[\operatorname{Pf}\widetilde{\mathcal H}_{\mathrm{lad}}(\pi)\bigr]\\
&=\operatorname{sgn}\!\bigl[(2t_1+\mu_1)(2t_2+\mu_2)-t_{\perp}^2\bigr]\\
&\qquad\quad\times\;\operatorname{sgn}\!\bigl[(2t_1-\mu_1)(2t_2-\mu_2)-t_{\perp}^2\bigr].
\end{split}
\end{align}

\section{Topological criterion for a tight binding model}\label{Topological criterion}

To intuitively understand the constraints of the topologically nontrivial phase in Eq.~\eqref{topological constraints0}, we rewrite these constraints as 
\begin{align}\label{topological constraints}
\begin{split}
t_{\perp}^2+\left[(t_2-t_1)-\frac{\mu_2-\mu_1}{2}\right]^2> \left[(t_1+t_2)-\frac{\mu_1+\mu_2}{2}\right]^2,\\
t_{\perp}^2+\left[(t_2-t_1)+\frac{\mu_2-\mu_1}{2}\right]^2< \left[(t_1+t_2)+\frac{\mu_1+\mu_2}{2}\right]^2.   
\end{split}
\end{align}
The topologically nontrivial phase is located between two cylindrical surfaces. To see this, we first perform a variable transformation: $\delta t=(t_2-t_1)/2$, $\delta\mu=(\mu_2-\mu_1)/2$, $t_{||}=(t_1+t_2)/2$ and $\mu_{||}=(\mu_1+\mu_2)/2$. Under this transformation, the topological constraints take the form 
\begin{align}\label{topological constraints2}
\begin{split}
t_{\perp}^2+\left(2\delta t-\delta\mu\right)^2> R_{1}^2,\\
t_{\perp}^2+\left(2\delta t+\delta\mu\right)^2< R_{2}^2,   
\end{split}
\end{align}
where $R_1=|2t_{||}-\mu_{||}|$ and $R_2=|2t_{||}+\mu_{||}|$. One can fix $2t_{||}\pm \mu_{||}$, since they are independent of other variables, so  $t_{\perp}^2 + \left(2\delta t \pm \delta\mu\right)^2= R_{1,2}^2$ represents two cylindrical surfaces whose axes lie along the directions $(\mp \delta \mu, 2\delta t)$. For $\delta t$ or  $\delta \mu=0$, the two cylindrical surfaces lie in the same direction. Note that since $R_1<R_2$,  the topologically nontrivial phase is confined between the two cylindrical surfaces. More rigorously, we note that 
\begin{align}
|2\delta t-\delta\mu|\leq 2|\delta t|+|\delta\mu|< R_{2} 
\end{align}
holds for $t_\alpha,\mu_\alpha>0$, by $|2\delta t|<2t_{||}$ and $|\delta\mu|<\mu_{||}$. This bound follows directly from the positivity assumptions and does not introduce a separate maximum parameter.

For $t_\perp=0$, the ladder separates into two independent Kitaev chains. The relative interleg superconducting phase can then be removed by independent gauge transformations on the two legs. In this decoupled limit, the total class-D invariant factorizes as
\begin{align}
\mathcal M=\mathcal M_1\mathcal M_2,
\qquad
\mathcal M_\alpha
=
\operatorname{sgn}
\left[(\mu_\alpha-2t_\alpha)(\mu_\alpha+2t_\alpha)\right].
\end{align}
Here $\mathcal M_\alpha=-1$ denotes an individually topological leg.

Three cases must be distinguished. If exactly one leg is topological, the total class-D parity is nontrivial, $\mathcal M=-1$. If both legs are topological, then the total class-D parity is trivial, $\mathcal M=+1$, although exact leg decoupling protects two Majorana zero modes at each boundary. If neither leg is topological, then $\mathcal M=+1$ and no Majorana end modes occur.

Thus, on the $t_\perp=0$ axis, the pair $(\mathcal M_1,\mathcal M_2)$ provides information that is not retained by the total class-D invariant alone. Equation \eqref{topological constraints0} at $t_\perp=0$ identifies the case in which exactly one leg is topological:
\begin{align}
(2t_1-\mu_1)(2t_2-\mu_2)<0,
\end{align}
under the positive-parameter assumptions used here. The region with
\begin{align}
(2t_1-\mu_1)(2t_2-\mu_2)>0
\end{align}
contains the two cases in which the legs share the same topology and must therefore be distinguished using the individual leg invariants.

If one of the two Kitaev chains is chosen to be at the topological boundary with $2t_\alpha=\mu_\alpha$, the condition for the ladder to be topologically nontrivial is given by 
\begin{align}
0<t_{\perp}^2<(2t_1+\mu_1)(2t_2+\mu_2).
\end{align} 
In this case the inner cylindrical boundary vanishes.

\section{Symmetry Analysis}\label{Symmetry Analysis}
We now turn to an analysis of the symmetry constraints for the Josephson diode effect with a very small or even vanishing barrier potential. Below, $\mathcal H(x,\phi,\theta)$ denotes the BdG operator in Eq.~\eqref{Ham-cont}, with the phase parameters shown explicitly. Since the dc Josephson CPR is given by 
\begin{align}\label{JCP}
I(\theta)=\frac{2e}{\hbar}\frac{\partial F}{\partial \theta}, \quad F=-\frac{1}{2} \sum_{E_{n}>0} E_{n} (\theta)
\end{align}
the existence of any symmetry transformation of the Hamiltonian from $\mathcal{H}(x,\theta)$ to $\mathcal{H}(x,-\theta+\tilde{\theta})$ (current reversal relation) suppresses the diode effect~\cite{Josephsontopology, Wang2025CurrentReversion}, where $\tilde{\theta}$ is a constant global phase. The existence of the diode effect therefore relies on the absence of all such current-reversing symmetries.

We analyze the spatial inversion $\mathcal P$, the effective antiunitary operation $\widetilde{\mathcal T}=\mathcal K$ with $\widetilde{\mathcal T}^{2}=+1$, and the leg-space pseudo-mirror operations $\mathcal M'_i=\tau_0\otimes\sigma_i$ $(i=x,y,z)$, with Nambu ordering $\Phi=(\psi_1,\psi_2,\psi_1^\dagger,\psi_2^\dagger)^T$. Here $\mathcal P$ acts as $x\mapsto -x$ (which also sends $p_x\mapsto -p_x$) and reverses the direction of the current, whereas $\mathcal M'_i$ act only in the leg pseudospin space and should not be confused with ordinary spatial mirror operations. The pseudo-mirror operations $\mathcal M'_x$ and $\mathcal M'_y$ flip the interleg superconducting phase as $\phi\to -\phi$ (up to a global phase of $2\pi$). First consider the case when $\delta\mu=0$ and $U(x)=0$, for which the Hamiltonian reduces to
\begin{align}
\begin{split}
\mathcal{H}(x)=&\Bigl[\Bigl(\frac{p_{x}^2}{2m}-\mu\Bigr)\sigma_{0}-t_{\perp}\cos\frac{\phi}{2}\,\sigma_{x}\Bigr]\tau_{z}\\
&-t_{\perp}\sin\frac{\phi}{2}\,\sigma_{y}\,\tau_{0}\\
&-\frac{\Delta}{2\hbar}\Bigl\{p_{x},\,\sin\theta(x)\,\tau_{x}-\cos\theta(x)\,\tau_{y}\Bigr\},
\end{split}
\end{align}
Under spatial inversion, $p_{x}\mapsto -p_{x}$ and $\theta(x)\mapsto\theta(-x)$. The $\tau_{z}$-diagonal part is even in $p_{x}$ and therefore unchanged. For the pairing term,
$\{p_{x},f(x)\}\mapsto\{-p_{x},f(-x)\}=-\{p_{x},f(-x)\}$. On each constant-$\theta$ side the sign reversal of the pairing, together with the left-right exchange $x\leftrightarrow -x$, maps the junction phase as $\theta\mapsto-\theta$ (the $p_{x}\mapsto -p_{x}$ sign is equivalent to a common gauge shift of both pairing phases and drops out of the phase difference). Consequently,
\begin{align}
\mathcal{P}\mathcal{H}(x,\phi,\theta)\mathcal{P}^{-1}\simeq\mathcal{H}(x,\phi,-\theta),
\end{align}
where $\simeq$ denotes equality up to a common superconducting gauge transformation, giving the current-reversal relation $I(\theta)=-I(-\theta)$ and prohibiting a diode response.

Under $\widetilde{\mathcal T}=\mathcal K$, the pairing term transforms as $\mathcal K\{p_{x},\sin\theta\,\tau_{x}-\cos\theta\,\tau_{y}\}\mathcal K^{-1}
=\{p_{x},\sin(-\theta)\,\tau_{x}-\cos(-\theta)\,\tau_{y}\}$ because $p_{x}$ and $\tau_{y}$ both change sign under complex conjugation ($\mathcal K p_{x}\mathcal K^{-1}=-p_{x}$, $\mathcal K\tau_{y}\mathcal K^{-1}=-\tau_{y}$, while $\mathcal K\tau_{x}\mathcal K^{-1}=\tau_{x}$), and the anticommutator with $p_{x}$ contributes one overall sign. The normal-state part transforms as $\mathcal K(\sigma_{y})\mathcal K^{-1}=-\sigma_{y}$, flipping the sign of the $\sigma_{y}\tau_{0}$ term. The combined result is
\begin{align}
\widetilde{\mathcal T}\mathcal{H}(x,\phi,\theta)\widetilde{\mathcal T}^{-1}=\mathcal{H}(x,-\phi,-\theta).
\end{align}
In the continuum gauge of Eq.~\eqref{Ham-cont}, the effective antiunitary constraint is implemented by $\mathcal K$ at $\phi=0$ and by $(\tau_0\otimes\sigma_z)\mathcal K$ at $\phi=\pi$ (modulo $2\pi$). Both operations square to $+1$. For generic $\phi\neq0,\pi$ these constraints are broken.

The leg-space pseudo-mirror operations act as ${\mathcal M'_i}\,\mathcal H\,{\mathcal M'_i}^{-1}$ with $\mathcal M'_i=\tau_0\otimes\sigma_i$. Using the algebra of the Pauli matrices one finds, for $\delta\mu=0$,
\begin{align}
\mathcal{M}_{x}'\mathcal{H}(x,\phi,\theta)\mathcal{M}_{x}'^{-1}&=\mathcal{H}(x,-\phi,\theta),\\
\mathcal{M}_{y}'\mathcal{H}(x,\phi,\theta)\mathcal{M}_{y}'^{-1}&=\mathcal{H}(x,-\phi+2\pi,\theta),\\
\mathcal{M}_{z}'\mathcal{H}(x,\phi,\theta)\mathcal{M}_{z}'^{-1}&=\mathcal{H}(x,\phi+2\pi,\theta).
\end{align}
The effective spinless time-reversal-like symmetry $\widetilde{\mathcal T}$ is broken if $\phi\notin\{0,\pi\}\pmod{2\pi}$. The pseudo-mirror operations $\mathcal M'_{x/y}$ reverse the interleg phase $\phi$ up to a shift by $2\pi$, whereas $\mathcal M'_z$ implements the gauge-equivalent change $\phi\to\phi+2\pi$. For $\delta\mu=0$, the two combined current-reversing operations have distinct parameter mappings:
\begin{align}
\widetilde{\mathcal T}\mathcal M'_x\,\mathcal H(x,\phi,\theta)\,
(\widetilde{\mathcal T}\mathcal M'_x)^{-1}
&=\mathcal H(x,\phi,-\theta),\\
\widetilde{\mathcal T}\mathcal M'_y\,\mathcal H(x,\phi,\theta)\,
(\widetilde{\mathcal T}\mathcal M'_y)^{-1}
&=\mathcal H(x,\phi-2\pi,-\theta).
\end{align}
The second right-hand side is unitarily equivalent to $\mathcal H(x,\phi,-\theta)$ through $\mathcal M'_z$. Thus both operations impose the same spectral relation and forbid a diode response when $\delta\mu=0$.

For the junction studied here, the spatially antisymmetric profile $\delta\mu(-x)=-\delta\mu(x)$ breaks these current-reversing constraints. Under inversion, the term $\delta\mu(x)\sigma_z\tau_z$ becomes $\delta\mu(-x)\sigma_z\tau_z=-\delta\mu(x)\sigma_z\tau_z$. Under either $\widetilde{\mathcal T}\mathcal M'_x$ or $\widetilde{\mathcal T}\mathcal M'_y$, the leg-space operation likewise reverses $\sigma_z$, while complex conjugation leaves it unchanged. Hence none of these transformations maps the Hamiltonian back to the same junction parameters for nonzero $\delta\mu(x)$. At generic $\phi$, this removes the current-reversing constraints identified above and permits a diode response when the relevant current contributions are sufficiently asymmetric.

\bibliography{Reference}
\end{document}